\documentclass[journal,comsoc]{IEEEtran}

  \usepackage[pdftex]{graphicx}

\usepackage{amsmath}
\usepackage{amsfonts,amssymb}
\usepackage{multirow}
\usepackage{color}
\usepackage{scalerel}

\usepackage{tikz}
\usetikzlibrary{tikzmark}
\usetikzlibrary{positioning}

\newtheorem{proposition}{Proposition}
\newtheorem{theorem}{Theorem}

\newtheorem{lemma}{Lemma}

\newcommand{\qed}{\hfill \IEEEQED}

\def\Label#1{\label{#1}\ [\ \text{#1}\ ]\ }
\def\Label{\label}

\begin{document}
\title{Covert communication with Gaussian noise: from random access channel to point-to-point channel\thanks{MH is supported in part by the National Natural Science Foundation of China (Grant No. 62171212) and
Guangdong Provincial Key Laboratory (Grant No. 2019B121203002).}}
 \author{
Masahito~Hayashi,~\IEEEmembership{Fellow,~IEEE} 
        and
\'{A}ngeles V\'{a}zquez-Castro,~\IEEEmembership{Senior Member,~IEEE}
\thanks{Masahito Hayashi is with
School of Data Science, The Chinese University of Hong Kong, Shenzhen, Longgang District, Shenzhen, 518172, China,
International Quantum Academy (SIQA), Futian District, Shenzhen 518048, China,
and
Graduate School of Mathematics, Nagoya University, Chikusa-ku, Nagoya 464-8602, Japan.
e-mail: hmasahito@cuhk.edu.cn, hayashi@iqasz.cn.
}
\thanks{
\'{A}ngeles V\'{a}zquez-Castro
is with the Department of Telecommunications and Systems Engineering, and with The Centre for Space Research (CERES) of Institut d'Estudis Espacials de Catalunya (IEEC-UAB) at 
Autonomous University of Barcelona,
Barcelona, Spain.
e-mail: angeles.vazquez@uab.es.}\thanks{Manuscript submitted 30th June 2020; revised xxx, 2020.}}
\markboth{M. Hayashi 
and \'{A}. V\'{a}zquez-Castro: Covert communication: from random access channel}{}

\maketitle

\begin{abstract}
We propose 
a covert communication protocol for the
spread-spectrum multiple
random access with additive white Gaussian noise (AWGN) channel.
No existing paper has studied covert communication for the random access channel.
Our protocol assumes binary discrete phase-shift keying (BPSK) modulation, and it works well
under imperfect channel state information (I-CSI) for both
the legitimate and adversary receivers,
which is a realistic assumption in the low power regime.
Also, our method assumes that 
the legitimate users share secret variables in a similar way as the preceding studies.
Although several studies investigated the covert communication for the point-to-point communication, 
no existing paper considers the covert communication 
under the above uncertainty assumption 
even for point-to-point communication.
Our protocol under the above uncertainty assumption allows $O(n)$ legitimate senders and $O(n/\log n)$ active legitimate senders.
Furthermore, our protocol can be converted to a protocol for point-to-point 
communication that works under the above uncertainty assumption. 
\end{abstract}

\begin{IEEEkeywords}
Covert communication,
Information hiding,
Additive white Gaussian noise,
Random access channel,
Central limit theorem,
Universal code
\end{IEEEkeywords}

\section{Introduction}
\subsection{Background: point-to-point covert communication}
Covert communication is a technology to hide the existence of communication,
and has been actively studied.
This type of communication is often called 
communication with low probability of detection.
In this technology, the legitimate sender intends to 
transmit an information message to the legitimate receiver
while making such communication undetectable by the adversary.
This task can be achieved 
when adversary's observation with the silent case
is imitated by adversary's observation under the existence of a communication 
between the legitimate sender and the legitimate receiver.
In fact, when the output of the silent case is written as 
a convex combination of other outputs in the channel to the adversary,
the above task can be easily achieved. 
Here, we call this condition the redundant condition,
and its rigorous definition is given in Section \ref{S3-25}.
Under the above condition, using the method of wire-tap channel \cite{Wyner,CK2,hay-wire},
the papers \cite{Hou,Wang} consider this problem for the point-to-point channel under the discrete memoryless condition.
They showed that the covert transmission length $O(n)$ is possible 
with $n$ uses of the channel.
The idea of this method is that
the transmitter transmits an independent and identically distributed (i.i.d.) sequence for the
no-communication mode, which makes the problem more similar to covert communication in
the presence of a jammer as discussed in 
\cite[Remark 2]{ZBN22}.

However, the redundant condition does not hold in general.
For example, when the form of the channel to the adversary is known,
the additive white Gaussian noise (AWGN) channel and 
the binary symmetric channel (BSC) do not satisfy this condition.
The papers \cite{Bash} and \cite{Che,CBCJ} discussed this problem 
for the cases of AWGN and BSC, respectively.
Then, the papers \cite{Wang,Bloch} studied the covert communication problem
under general discrete memoryless channel for the point-to-point channel
when the redundant condition does not hold.
They showed that the optimal covert transmission length is $O(\sqrt{n})$ 
in the non-redundant case with $n$ uses of the channel.
Following these studies, the papers \cite{Yan,Wang2,Zhang} studied this problem for AGWN as considering continuous time models.
Also, the paper \cite{Abdelaziz} extended this discussion to 
multiple-input multipl-output (MIMO) AWGN channels.

The fundamental assumption of
the non-redundant case is using preshared secure keys between 
the legitimate sender and the legitimate receiver.
Due to the preshared keys, 
the legitimate users can realize covert communication
even when the channel to the adversary has smaller noise than 
the channel to the legitimate receiver.
That is, to preshare keys is a mandatory resource for covert communication whenever Willie's channel is not worse than Bob's channel.
Later, many subsequent studies \cite{Lee,He,Sobers,Soltani,Goeckel,Hu,Shmuel,Xiong} 
analyzed the point-to-point covert communication over the AWGN channel
under different additional assumptions.

\begin{figure*}
\begin{center}
\includegraphics[scale=0.6]{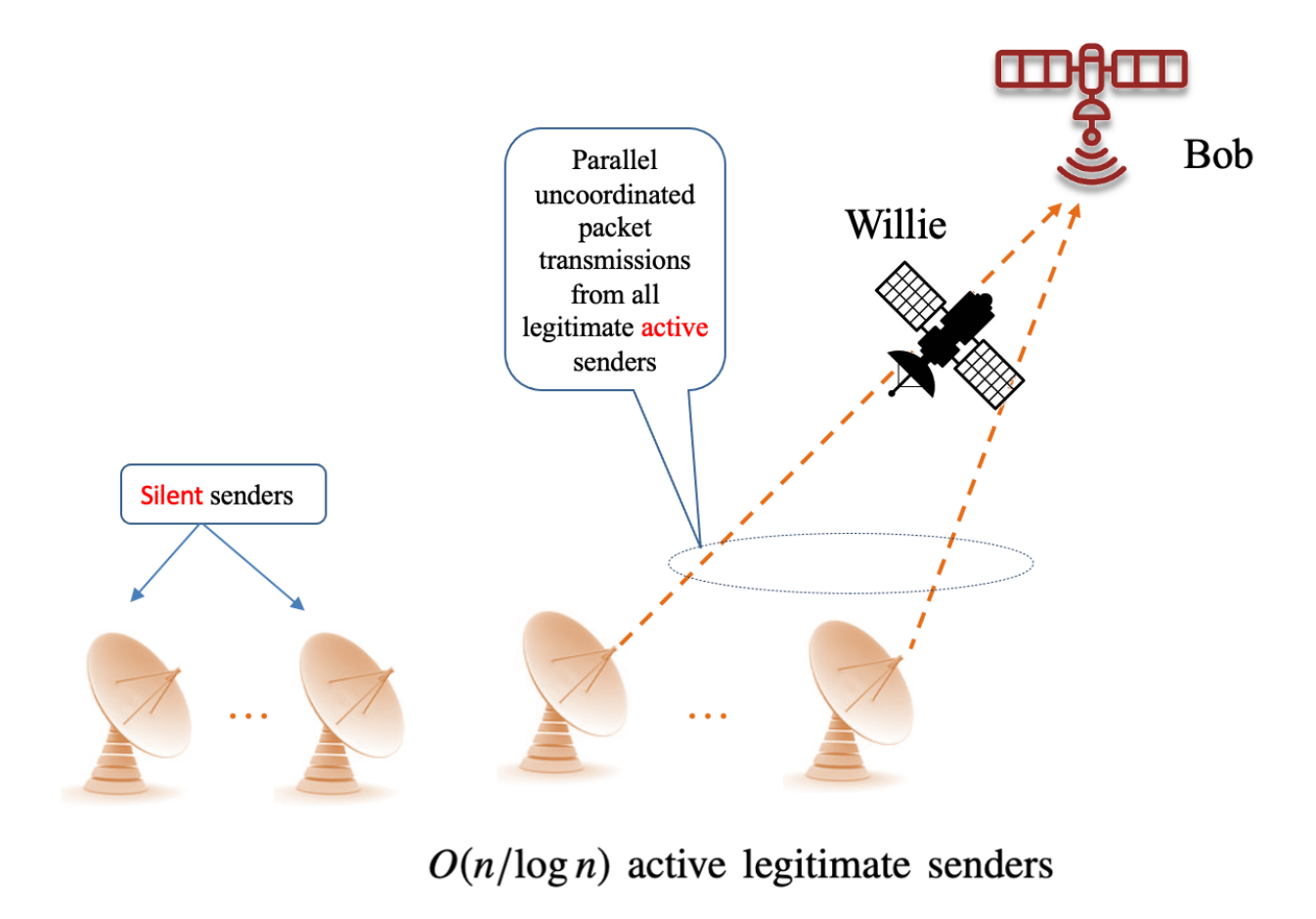}  
\end{center}
\caption{This figure illustrates a random access channel with one legitimate receiver (Bob),
one adversary (Willie),
and multiple legitimate senders. Our protocol allows 
$O(n/ \log n)$ active legitimate senders. 
Illegitimate receivers are shown in black.}
\Label{FF3C}
\end{figure*}

\subsection{Background: random access channel}
Our focus in this work is the random access channel where the point-to-point is a special case. A number of studies \cite{Minero,Yavas,Yavas2,Ordentlich,Ebrahimi} have already discussed the random access channel
from the viewpoint of information theory, however, these works do not address the covertness property of this channel.
Recently, the paper \cite{AV} addressed 
the anti-jamming secrecy
in the random access channel,
but it did not discuss covertness.
Another recent paper \cite{Arumugam} discussed the covert communication 
for the access channel assuming discrete time, 
but it did not address random access channel nor AWGN channel.
\subsection{Problem statement and novelty}
In this paper, we present an information theoretical study of the covertness of 
(direct sequence) random access and one time pad encryption. We assume BPSK modulation and novel assumptions on the channel knowledge by the legitimate users and 
the adversary, Willie.
Furthermore, to realize covertness, 
depending on the legitimate sender, our method uses
preshared secret binary symbols, which is often called secret chips. Fig. \ref{FF3C} shows its illustrative scenario and the relation with existing results for access channel are summarized as Table \ref{com1} clarifying the novelty of our work. 
Throughout this paper, we assume logarithms with base $e$.

\begin{table}[t]
\caption{Comparison with existing results for access channel}
\label{com1}
\begin{center}
\begin{tabular}{|c|c|c|c|}
\hline
&\multirow{2}{*}{Covertness}& Random access & Type of \\
& &channel & channel \\
\hline
\cite{Arumugam} & Yes & No & Discrete \\
\hline
\cite{Minero,Yavas,Ebrahimi} & No & Yes  & Discrete\\
\hline
\cite{Yavas2,Ordentlich,AV} & No & Yes  & AWGN\\
\hline
This paper & Yes & Yes & AWGN \\
\hline
\end{tabular}
\end{center}
\end{table}

In our setting, 
$n$ expresses the number of uses of the channel during one coding block-length. 
Our method allows $O(n)$ legitimate senders and $O(n/\log n)$ active legitimate senders
under the assumption of preshared secure keys between 
these legitimate senders and the legitimate receiver\footnote{
In contrast, the recent paper \cite{Arumugam} considers the case when 
the number of senders is fixed to $K$ and the size of transmitted bits behaves as $o(\sqrt{n})$.}.
To achieve this performance, 
we 
pose a novel and realistic channel condition (in the low power regime) as follows
because it is impossible to achieve 
covertness of $O(n/\log n)$ bits 
in the non-redundant case
when the channel parameter is completely known to the adversary, i.e., 
the channel is identified by the adversary.
In a realistic scenario, it is difficult for 
both legitimate and adversary parties
to obtain a complete knowledge of the channel parameters.
Therefore, 
it is natural to assume that all parties (legitimate and non-legitimate) do not have a complete knowledge of the channel parameters
while we assume that the channel parameters are fixed during one coding block-length.
The latter assumption is justified 
when our signal model and asymptotic results hold within the channel coherence time.
Under the above realistic conditions for the uncertainty, 
our protocol guarantees that the legitimate receiver retrieves the message.

Another novelty in our formulation is 
the universality of our proposed codes.
In information theory a code is called a universal code when 
the code does not depend on the channel parameter in the above way, 
i.e., our code construction does not
require full knowledge of the channel \cite{CK,Uni-cont}.
In our case, this means that our method allows the dispersion of signal intensities from the senders in the detection of both receivers due to the effect of the fading fluctuation.
As a consequence, we make the realistic assumption that for a known scenario of interest, an upper and lower bound of the channel coefficients can be estimated.

In addition, our method assumes preshared secure keys between 
the legitimate sender and the legitimate receiver
in the same way as \cite{Wang,Bloch,Yan,Wang2,Zhang} and 
it works even when the channel to the adversary has smaller noise than 
the channel to the legitimate receiver.
That is,  for every coding block, each legitimate sender shares 
secret binary symbols
as pre-shared secrets with the legitimate receiver
while the legitimate sender can send only one bit. 
Hence, when the senders need to transmit $\ell$ bits, it is sufficient that
they repeat this protocol $\ell$ times.
In practice, this is realistically achieved using well known and widely available spreading codes. 
Each legitimate sender has $n$ different preshared secret binary symbols.
That is, the number of secret binary symbols at the legitimate receiver is $n$ times the number of legitimate senders.

Our encoder is very simple for random access channel.
That is, when a legitimate sender is active and intends to transmit one bit $X$,
the sender encodes the intended bit $X$ into $n$ channel inputs by using one time pad encryption with $n$ preshared secret binary symbols.
Then, the legitimate receiver recovers the transmitted bits by using the preshared secret bits.

The most novel point of our work is the covertness analysis for the adversary.
In our analysis, the covertness evaluation is reduced to 
the difference between 
the Gaussian distribution and the distribution of the weighted sample mean of $n$ independent random variables subject to 
the average output distribution of BPSK modulation.
Although a variant of the central limit theorem \cite{non-IID} guarantees that
the distribution of the weighted sample mean of $n$ independent random variables
approaches to a Gaussian distribution,
our covertness analysis needs the evaluation of the variational distance between 
the above two distributions.
When the fading coefficients from a sender in Willie's detection 
does not depend on the sender,
it is sufficient to discuss the variational distance between 
the distribution of the sample mean and a Gaussian distribution.
Such a case was discussed in \cite[(1.3)]{IID}.
However, the general case requires more difficult analysis.
Fortunately, the recent papers \cite{gap,non-IID2} have studied this mathematical problem
by using Poincar\'{e} constant \cite{gap,non-IID2,Bobkov,Bobkov2}.
Applying this result, we derive our covertness analysis.

\subsection{From random access channel protocol to point-to-point channel}
Another novelty of our work is that we consider the fact that our protocol can be
converted to a protocol for point-to-point communication that works under the above assumptions.
Under this conversion, we obtain
a covert communication protocol for the point-to-point channel
that has $ \lceil (l_n +1)/2\rceil$ different values as the channel input power,
and achieves the covert transmission of $O(n/\log n)$ bits,
where $l_n$ is the number of bits the legitimate sender wants to transmit to the legitimate receiver.
Although existing studies assuming the redundant case \cite{Hou,Wang} achieve covert transmission length $O(n)$ 
with $n$ uses of the channel,
existing studies assuming the non-redundant case \cite{Wang,Bloch,Yan,Wang2,Zhang}
achieve covert transmission length $O(\sqrt{n})$, which is much smaller than the transmission length of the conventional communication.
To resolve this problem, 
many subsequent researchers \cite{Lee,He,Sobers,Soltani,Goeckel,Hu,Shmuel,Xiong} 
 introduced 
the uncertainty of the channel parameters only of the channel to the adversary
under the AWGN channel.
In fact, due to the uncertainty, the adversary cannot distinguish 
the output of the Gaussian mixture input distribution from the output of 
zero input. 
That is, this modification enables the channel model to satisfy the redundant condition, which leads the covert transmission length $O(n)$.
However, these studies assume that the legitimate receiver knows the channel parameters of his/her own channel, which is an unequal assumption, i.e.,
an unrealistic assumption.
To make a fair assumption, we pose the novel and realistic channel condition introduced above
that the channel parameters of the AWGN channels to both the legitimate and adversary
receivers present some uncertainty, i.e. are not completely known to them.
Fortunately, our protocol on the point-to-point channel
achieves the covert transmission of $O(n/\log n)$ bits
when both receivers (the legitimate receiver and the adversary)
have uncertainty in their detection.
In this sense, our method has an advantage over existing methods
even under the Gaussian point-to-point channel.

Finally, we remark that our main focus is the analysis of the asymptotic performance of our code to guarantee covertness and therefore practical issues 
such as BPSK symbol acquisition or outage probability (e.g for concrete statistics assumptions on the channel dynamics) 
and practical channel estimation
are out of the scope of this work and is left for future work.
The relation with existing results for point-to-point channel
is summarized in Table \ref{com2}, clarifying the novelty of our work and results.

\begin{table}[t]
\caption{Comparison with existing results for point-to-point channel}
\label{com2}
\begin{center}
\begin{tabular}{|c|c|c|c|c|}
\hline
&\multirow{3}{*}{Transmission} & Bob & Willie & \multirow{2}{*}{Type} \\
&\multirow{3}{*}{length}& knows & knows & \multirow{2}{*}{of} \\
& &channel  &channel  & 
\multirow{2}{*}{channel} \\
&&parameters & parameters &\\
\hline
\cite{Bloch,Wang} &$O(\sqrt{n})$& Yes & Yes & Discrete \\
\hline
[11--13]
&$O(\sqrt{n})$& Yes & Yes & AWGN \\
\hline
[15--22]
&$O(n)$& Yes & No  & AWGN\\
\hline
This paper &$O(n/\log n)$& No & No & AWGN \\
\hline
\end{tabular}
\end{center}
$n$ is the number of uses of channel.
\end{table}

This paper is organized as follows.
Section \ref{S3} describes our formulation of random access channel model,
and states our result in this case.
Section \ref{S2} explains what protocol is obtained
for the particular case of point-to-point channel model.
Then, Section \ref{S2} compares our obtained code for 
the point-to-point channel model with simple applications of 
the methods \cite{Hou,Wang}.
Section \ref{S5} shows that Bob correctly recovers the message with almost probability one
under both models in the asymptotic case.
Section \ref{S5-2} formulates covertness with respect to Willie,
and states our covertness result. In addition, Section \ref{S5-2} shows its proof 
for the case with equal fading including the point-to-point channel model
while its proof with the general case with unequal fading is shown in Appendix.
Section \ref{S6} presents a discussion of our results.

\section{Random access channel}\Label{S3}
\subsection{Random access channel model}
Our random access channel model 
has $m$ senders ${\cal A}_1, \ldots, {\cal A}_m$,
one adversary, Willie, and the legitimate receiver, Bob.
The task of our protocol is formulated as follows.
Each sender ${\cal A}_i$ intends to send one bit $M_i$ to Bob
within the channel coherence time
when ${\cal A}_i$ is active.
If ${\cal A}_i$ is silent, he/she does not need to send it to Bob.
Also, the senders want to hide the existence of their communication to 
the adversary, Willie.
For the practical implementation, we assume that 
the channel is AWGN and each sender can use only BPSK modulation.

To realize the hidden communication, 
the senders and Bob share secret random variables that are not known to Willie, 
in the same way as \cite{Wang,Bloch}.
That is,
the sender ${\cal A}_i$ has binary random variables $S_{i,1}, \ldots, S_{i,n}$ that are subject to the uniform distribution independently.
The legitimate receiver, Bob also knows all the binary random $S_{i,j}$.
However the adversary, Willie, does not know $S_{i,j}$.

We consider a random access channel with Gaussian channel as follows.
Assume that 
only $l$ senders ${\cal A}_{i_1},\ldots, {\cal A}_{i_l}$ are active and other senders are silent.
When ${\cal A}_{i_k}$ inputs $n$ variable $X_{i_k,1}, \ldots, X_{i_k,n}$,
Bob receives
\begin{align}
Y_j= {N}_{B,j}+\sum_{k=1}^l a_{i_k} X_{i_k,j} \Label{MMAP}
\end{align}
for $j=1, \ldots, n$.
Similarly, Willie receives 
\begin{align}
Z_j=N_{W,j}+\sum_{k=1}^l b_{i_k} X_{i_k,j}
\Label{MMAP2}
\end{align}
for $j=1, \ldots, n$.
Here, $a_{i_k}$ and $b_{i_k}$ are the fading coefficients within the channel coherence time in Bob's and Willie's detection.
Hence, $a_{i_k}$ and $b_{i_k}$ are positive constants during a coherent time.
That is, we treat $a_{i_k}$ and $b_{i_k}$ as constants in the following discussion.
In the following, we denote the maximum $\max_{k} a_{i_k}$
($\max_{k} b_{i_k}$) and 
the minimum $\min_{k} a_{i_k}$ ($\min_{k} b_{i_k}$) 
by $\overline{a}$ ($\overline{b}$) and 
$\underline{a}$ ($\underline{b}$), respectively.
Hence, we make the realistic assumption that for a known scenario of interest, 
upper and lower bounds of the channel coefficients can be estimated.
That is, Bob and Willie know $\overline{a}$, $\overline{b}$, $\underline{a}$,
and $\underline{b}$.
Also, $N_{B,1}, \ldots, N_{B,n}$ ($N_{W,1}, \ldots, N_{W,n}$) 
are independent Gaussian variable with average $0$ and variance $v_B$ ($v_W$).

In a realistic setting, Bob and Willie know rough values of the variance of noise power for Bob and Willie, denoted as $v_B$ and $v_W$,
but, they do not know their exact values due to the following reason.
Willie and Bob know their own receiving device well.
Hence, they know the noise generated in their receiving device.
In this way, Bob and Willie have a similar performance for receiving the signal from the senders.
However, a part of the noise is generated out of Willie's device, 
which can be considered as a background noise.
To discuss Willie's detection of the existence of the communication, 
we focus on Willie's knowledge on the value of the variance $v_W$
of Willie's observation. 
That is,
we denote the set of possible variance $v_W$ of Willie's observation
by ${\cal V}$. 
In the following, we assume that ${\cal V}$ is an open set of $\mathbb{R}$.
This background noise uncertainty also affects to Bob, however, he has the pre-shared keys so that he can recover the message nevertheless such uncertainty.
In contrast, since Willie doesn't have it, 
his ability is affected by such background noise uncertainty.

In fact, it is a common assumption that the channel is characterized by unknown parameters. 
In this case, the channel model is usually denoted as compound, 
and codes for such channels are called universal codes in information theory \cite{CK}
(see e.g. the paper \cite{Uni-cont} for a Gaussian channel).
Even in the above existing universal setting,
we need to assume that 
the channel parameters belong to a certain subset.
Otherwise, it is impossible to guarantee secure communication.
Estimating channel coefficients is well known in signal processing. 
The transmission inserts ``pilots'' i.e. known symbols to measure the channel effect at reception. Hence, to assume some underlying ``roughly'' estimation of the channel dynamics is reasonable \cite{Sun}.
In the following, we propose our code that 
does not depend on these channel parameters except for $\underline{a}$
and $\frac{ v_B}{ \underline{a}^2}$.

\subsection{Random access covert protocol}
\Label{S-pro}
Here, we present our protocol for 
random access covert communication.
When the sender ${\cal A}_i$ is silent,
the input signal $X_{i,j}$ is set to zero for $j=1, \ldots, n$.
When the sender ${\cal A}_i$ is active and 
the sender ${\cal A}_i$ sends the binary message $M_i$,
${\cal A}_i$ encodes the message as
\begin{align}
X_{i,j}=\sqrt{\frac{t_n}{n}} (-1)^{M_i+S_{i,j}} \Label{NJ}
\end{align}
 for $j=1, \ldots, n$.
Here, the above code uses average power $\frac{t_n}{n}$
for each channel use, which is sufficiently small.

\begin{figure*}
\begin{center}
  \includegraphics[width=1\linewidth]{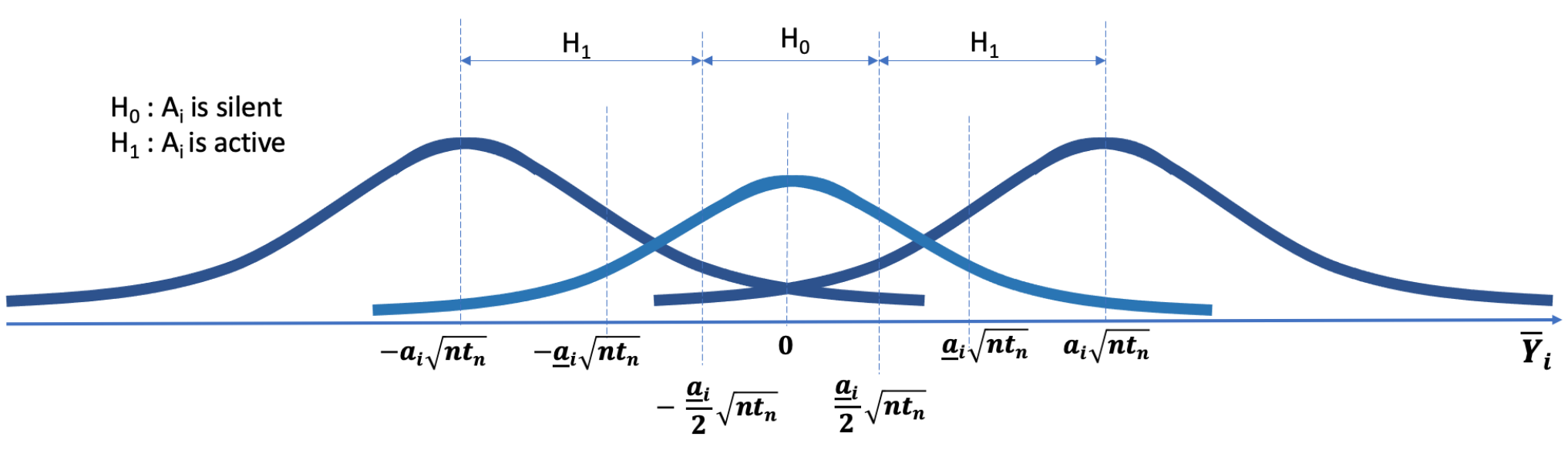}
  \end{center}
\caption{This figure shows how Bob makes his decision from $\bar{Y}_i$
for the random access case
when they use the channel $n$ times.
Here, Bob needs to output one of three outcomes, silent, $1$, and $0$.
The right graph shows the distribution of $\bar{Y}_i$ when $M_i=0$,
and the left graph shows the distribution of $\bar{Y}_i$ when $M_i=1$.
The central graph shows the distribution of $\bar{Y}_i$ when ${\cal A}_i$ is silent.
}
\Label{FF1C}
\end{figure*}

Next, we consider Bob's decoder. 
In order to recover $M_i$,
using the secret variables $S_{i,j}$, 
Bob calculates the decision statistic $\bar{Y}_i:= \sum_{j=1}^n (-1)^{S_{i,j}} Y_j$
from his receiving variables $Y_1, \ldots, Y_n $ as Fig. \ref{FF1C}.
For each $i$, Bob outputs one of three outcomes, silent, $1$, and $0$ as follows.
When $ |\bar{Y}_i| < \sqrt{n}\frac{\underline{a}t_n}{2}$,
Bob considers that ${\cal A}_i$ is silent.
When $ \bar{Y}_i \le -\sqrt{n}\frac{\underline{a}t_n}{2}$,
Bob considers that $M_i$ is 1.
When $ \bar{Y}_i \ge \sqrt{n}\frac{\underline{a}t_n}{2}$,
Bob considers that $M_i$ is 0.

Then, we have the following theorem for the analysis on the asymptotic performance of our code.
\begin{theorem}\Label{TH2}
Assume that the number of senders, $m$, and of active senders, $l$, 
and $m$, $l$ are given as $m_n$, $l_n=\frac{n}{t_n}$ to satisfy 
\begin{align}
\frac{n}{l_n^2}
 & \to 0 \Label{BZI2} ,\\
\frac{l_n}{n}
 & \to 0 \Label{NCA} ,\\
 \frac{n \underline{a}^2}{8l_n (   v_B+ 
  \overline{a}^2 )}
- \log m_n &\to + \infty \Label{BZI} .
\end{align}
Also, we assume that $v_W$ belongs to an open set ${\cal V}$.
Then, under the above presented protocol, 
Bob can recover the message with asymptotically zero error,
and Willie cannot detect the existence of the communication
regardless of the values of the channel parameter to Willie.
\end{theorem}
Here, we have not formulated Willie's detection.
In Section \ref{S5-2A}, we state the impossibility of
Willie's detection after presenting 
its formal definition.

Interestingly, our encoder and our decoder do not depend on 
the values $a_i,b_i,v_B,v_W$ of the channel parameters,
and our decoder do not depend on
the number $l_n$ of active senders.
But, the probability of correct decoding 
depends on the number $l_n$, and is close to $1$
as long as 
 the conditions \eqref{BZI2} and \eqref{BZI} 
hold.
Hence, in order that Bob knows whether his decoding is correct,
he needs to know whether
the number $l_n$ is smaller than a certain threshold, which can deduced by Bob from the estimated received power.

For example, 
when 
$l_n= c n \log n$ with 
$c< \frac{8 (v_B+ \overline{a}^2 )}{\underline{a}^2}
$
and $m_n=O(n)$,
the conditions \eqref{BZI2} and \eqref{BZI} hold.
Also, 
when 
$l_n= c n \log n$ with 
$c\le \frac{8 (v_B+ \overline{a}^2 )}{\underline{a}^2}
$
and $m_n=l_n$,
these conditions hold.
Therefore, 
covert communication with random access code
is asymptotically possible with 
$O(n)$ senders and $O(n/\log n)$ active senders
when all active senders transmit only one bit.

Now, we consider the case when each active sender ${\cal A}_i$ 
wants transmit $u_n=g(n)$ bits, $M_{i,1}, \ldots, M_{i,u_n}$.
In this case, 
the active sender ${\cal A}_i$ shares random binary symbols
$S_{i,1,j}, \ldots, S_{i,u_n,j}$ with Bob for $j=1, \ldots, n$, 
and
the active sender ${\cal A}_i$ 
sets $X_{i,j}$ as
\begin{align}
X_{i,j}= \sqrt{\frac{t_n}{n}} \sum_{j'=1}^{u_n}
(-1)^{M_{i,j'}+ S_{i,j',j}}
\end{align}
as the encoding.
Here, we denote the numbers of senders and active senders
by $m_n'$ and $l_n'$.
This situation can be considered as a special case of 
Theorem \ref{TH2} 
with $m_n= g(n)m_n' $ and $l_n=g(n) l_n'$.
Hence, when
$m_n= g(n)m_n' $ and $l_n=g(n) l_n'$
satisfy the conditions \eqref{BZI2} and \eqref{BZI},
Bob can recover the message with asymptotically zero error,
and Willie cannot detect the existence of the communication
regardless of the values of the channel parameter to Willie.

Also, the condition \eqref{BZI} implies that $t_n$ goes to zero.
Since $t_n$ is the power per user,
the power per user
needs to be zero asymptotically in our protocol.
This agrees with the intuition that covertness requires as low power as possible.
Further, since the values of the channel parameter to Willie are not contained in the assumption 
of this theorem, the covert communication is possible 
even if Willie's channel is better than Bob's channel.

\begin{figure*}
\begin{center}
  \includegraphics[width=1\linewidth]{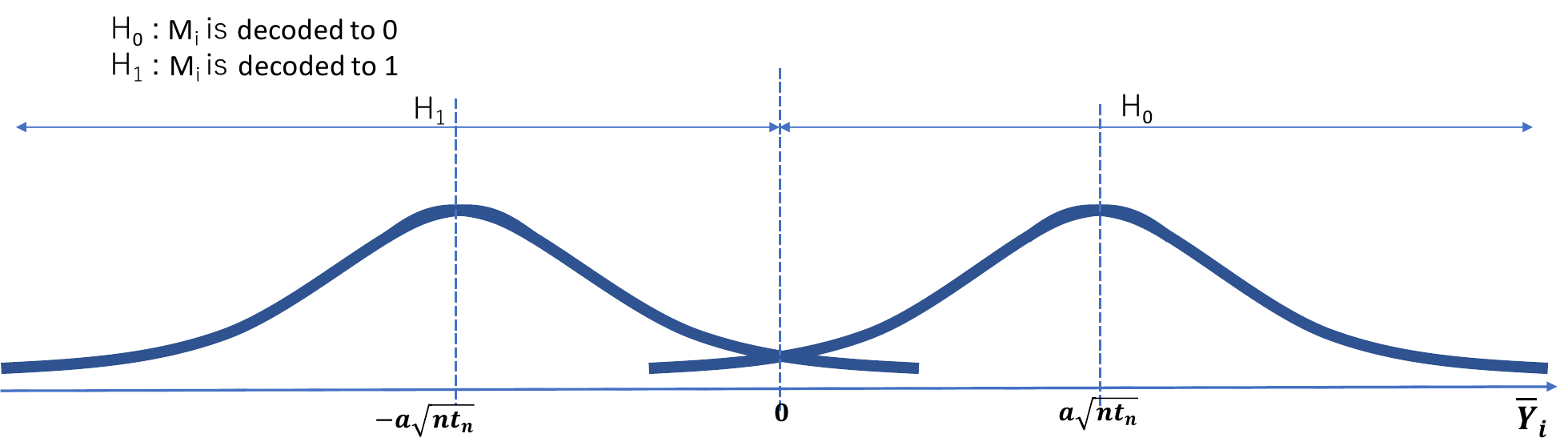}
  \end{center}
\caption{This figure shows how Bob decodes the message $M_i$ from $\bar{Y}_i$
for the point-to-point case.
The right graph shows the distribution of $\bar{Y}_i$ when $M_i=0$,
and the left graph shows the distribution of $\bar{Y}_i$ when $M_i=1$.
}
\Label{FF2C}
\end{figure*}

\section{Point-to-point channel}\Label{S2}
\subsection{Point-to-point channel model}
In the following, we discuss what protocol is obtained when 
the above protocol is applied to the point-to-point channel model.
In this model,
the legitimate sender, Alice, intends to transmit $l_n$ bits to the legitimate receiver, Bob,
with $n$ uses of AWGN channel 
while the intensity of input can be fixed to a single value intended by Alice
during one block length.
This setting is often called point-to-point communication.
Table \ref{term} summarizes the relation between 
random access channel and point-to-point channel.

\begin{table}[t]
\caption{Relation between
random access channel and point-to-point channel}
\label{term}
\begin{center}
\begin{tabular}{|c|c|c|}
\hline
& random access & point-to-point   \\
&channel  &channel  \\
\hline
$n$ & number of uses of channel & block length  \\
\hline
\multirow{2}{*}{$l_n$} & transmission length & transmission   \\
 & (number of active senders) &  length \\
\hline
$m_n$ & number of senders & this number is set to be $l_n$ \\
\hline
\end{tabular}
\end{center}
\end{table}

When Alice's $j$-th input variable is $X$, 
Bob receives
\begin{align}
Y_j= {N}_{B,j}+ a X_{j} \Label{MMAP-A}
\end{align}
for $j=1, \ldots, n$.
The variables ${N}_{B,1}, \ldots, N_{B,n}$ are independent Gaussian random variables
with mean $0$ and variance $v_B$. 
Similarly, Willie receives 
\begin{align}
Z_j=N_{W,j}+b X_{j}
\Label{MMAP2-A}
\end{align}
for $j=1, \ldots, n$.
The variables ${N}_{W,1}, \ldots, N_{W,n}$ are independent Gaussian random variables
with mean $0$ and variance $v_W$. 
Then, we make the same assumption for $v_B$  and $v_W$ as the previous section.
Here, when Alice is silent, all input variables $X_1, \ldots, X_n$ are zero.
Since $a$ and $b$ are the fading coefficients in Bob's and Willie's detection,
$a$ and $b$ are positive constants during a coherent time.
That is, we treat $a$ and $b$ as constants in the following discussion.
Therefore, our model is the model in the previous section with 
$ \underline{a}=\overline{a}=a$, $ \underline{b}=\overline{b}=b$, and $m=l$.


\subsection{Covert protocol}
For covert transmission of $l_n$ bits $M_1, \ldots, M_{l_n} $ within the coherence channel time,
Alice and Bob share $n l_n$ secret binary symbols
$\{S_{i,j}\}_{1 \le j \le n, 1 \le i \le l_n} $.
Then, Alice encodes the message $M_1, \ldots, M_{l_n} $ as 
\begin{align}
X_{j}=\sqrt{\frac{t_n}{n}} \sum_{i=1}^{l_n} (-1)^{M_i+S_{i,j}} \Label{NJ-A}
\end{align}
 for $j=1, \ldots, n$.
Here, the above code uses average power $\frac{t_n}{n}$
for each channel use, which is sufficiently small.

To see another form of $X_{j}$, we define $A_{j}$ as 
$A_{j}:= |\{ i| M_{i}+S_{i,j}=1 \hbox{ mod } 2\}|$,
where $|A|$ expresses the number of elements of the set $A$.
Then, $A_{j}$ is independently subject to 
the binary distribution with 
$l$ trials and with probability $1/2$.
$X_j$ has another form as
\begin{align}
X_j=\sqrt{\frac{t_n}{n}} (l_n-2A_j).
\end{align}
The variable $\sqrt{\frac{t_n}{n}}( l_n-2 A_{j} )$ has average $0$
and variance $ \frac{t_n l_n}{n} $.
Hence, the power for one channel use, i.e., the expectation of $X_j^2$ is $ \frac{t_n l_n}{n} $, which converges to zero.

Next, we consider Bob's decoder. 
In order to recover $M_i$,
using the secret variables $S_{i,j}$, 
Bob calculates the decision statistic $\bar{Y}_i:= \sum_{j=1}^n (-1)^{S_{i,j}} Y_j$
from his receiving variables $Y_1, \ldots, Y_n $
as Fig. \ref{FF2C}.
When $ \bar{Y}_i \le 0$, Bob considers that $M_i$ is 1.
When $ \bar{Y}_i > 0$, Bob considers that $M_i$ is 0.
Then, we have the following theorem for the analysis on the asymptotic performance of our code.
Hence, as a special case of Theorem \ref{TH2}, we have 
the following theorem.

\begin{theorem}\Label{TH1}
When $v_W$ belongs to an open set ${\cal V}$
and $l=l_n$ satisfies
\eqref{BZI2} and 
\begin{align}
 \frac{n {a}^2}{8l_n (   v_B+ 
  {a}^2 )}
- \log l_n &\to + \infty ,
\Label{ALZ2-A}
\end{align}
Bob can recover the message with asymptotically zero error,
and Willie cannot detect the existence of the communication.
\end{theorem}

Similar to Theorem \ref{TH2}, 
we have not formulated Willie's detection.
In Section \ref{S5-2A}, we state the impossibility of
Willie's detection after presenting 
its formal definition.

This theorem shows that
covert communication is asymptotically possible with transmission length
$c n \log n$ with 
$c\le \frac{8 (v_B+ \overline{a}^2 )}{\underline{a}^2}
$.
Interestingly, our encoder and our decoder does not depend on 
the values $a,b,v_B,v_W$ of the channel parameters 
 as long as the condition \eqref{ALZ2-A} holds.

\subsection{Relation with redundant condition}\Label{S3-25}
The existing studies \cite{Hou,Wang}
showed the following.
When the channel to Willie
satisfies the redundant condition,
a better transmission rate.
However, our channel model does satisfy this condition even when the channel parameters of the channel to Willie 
are fixed.
Therefore, these existing results cannot be applied to our case.
To see this fact,
we recall the definition of 
the redundant condition.

We denote the output sample space in the channel to Willie
by ${\cal Z}$,
which is potentially an infinite set.
We also denote a measure on ${\cal Z}$ by $\nu(dz)$.
We denote the set of Alice's input by ${\cal X}$,
which is a finite set.
Depending on Alice's input $x \in {\cal X}$,
Willie's output is subject to the probability density function
$p_{Z,x} $.
When the following condition holds,
the channel to Willie is called redundant.
There exist two non-identical distributions
$P_X$ and $P_X'$ such that
\begin{align}
\sum_{x \in {\cal X}}P_X(x)p_{Z,x}=
\sum_{x \in {\cal X}}P_X'(x)p_{Z,x}.
\end{align}

In our model \eqref{MMAP2-A}, 
the Willie's output is simplified as
\begin{align}
Z=N_{W}+b X. 
\end{align}
Here, the variable ${N}_{W}$ is a Gaussian random variables
with mean $0$ and variance $v_W$. 
Since 
our channel model of the channel to Willie 
does not satisfy 
the redundant condition with given $v_W$ and $b$,
we cannot directly apply the existing method by \cite{Hou,Wang} to our model.

\subsection{Comparison with existing methods}\Label{S3-3}
Many recent researchers \cite{Lee,He,Sobers,Soltani,Goeckel,Hu,Shmuel,Xiong} 
have obtained achievability of covert transmission of
$O(n)$ bits when only Willie's channel has uncertainty and 
Bob knows a certain knowledge of his own channel unlike our assumption.
When the fading coefficient $a$ is known and only the noise power $v_B$ is unknown,
the decoder for the case with the maximum noise power works well.
Hence, the existing method does work well in this case.
For example, the paper \cite{Sobers} assumes that the fading coefficient $a$ is 1
while  $v_B$ is unknown\footnote{The paper \cite{CBCJ} considers the case when the channel parameter is unknown. But, it assumes the BSC channel, which is different from AWGN channel.}.
However, when the fading coefficient $a$ is unknown in addition to $v_B$,
the conventional decoder does not work because 
the decoder depends on the value of the fading coefficient $a$.
To understand this difficulty, consider the following case with $n=1$.
The sender uses four points $\pm x_1$ and $\pm x_0$ in $\mathbb{R}$ 
with $x_1>x_0>0$ for the encoding.
When the receiver receives a value $y \in (x_0,x_1)  \subset \mathbb{R}$,
the maximum likelihood (ML) decoder depends on the value of the fading coefficient $a$.
When $a< \frac{2y}{x_1-x_0} $,
the ML decoder estimates that the input is $+x_1$.
Otherwise, it estimates that the input is $+x_0$.
In this way, the decoder depends on the value of the fading coefficient $a$ in general.
In the case of conventional channel coding,
people often employ universal coding to resolve this problem.
The method of type \cite{CK} enables us to develop universal coding for the discrete memoryless channel case.
The paper \cite{Uni-cont} proposed a universal coding that works 
for the continuous case including the AWGN channels.
In this way, it is a crucial technology to develop a code that works independently of the channel parameters.

To see the difficulty to achieve the covert transmission
under imperfect knowledge for both channels,
the following part discusses what problem appears in simple applications of the original methods \cite{Wang,Hou} under our setting.
Our channel model has a uncertainty for the variance $v_W$ 
of the additive Gaussian noise in the channel to the adversary.
This situation is formulated as follows.
Users know that
the variance of 
the additive Gaussian noise in the channel to the adversary,
Willie, belongs to an open set ${\cal V}$,
but they cannot identify 
which value in ${\cal V}$ is the true value $v_W$. 

Since Alice is allowed to use various values as the channel input power in 
the Gaussian point-to-point channel,
Alice can select the input $X_j$ freely.
Hence, we can apply the method \cite{Wang} for the redundant case
as follows.
First we choose a sufficiently small positive number $v'$ such that $v_W+b^2 v'$ belongs to ${\cal V}$.
Here, $v'$ expresses the possible 
error for Willie's knowledge about 
the variance of 
the additive Gaussian noise in the channel to Willie.
That is, even when the true value is $v_W$,
Willie cannot identify which of $v_W$ and 
$v_W+b^2 v'$ is the true.
Hence, Willie has to keep the possibility that 
$v_W+b^2 v'$ is the true as well as $v_W$.

Let $R_{v'}$ be a real number such that $R_{v'}< I(Y_j;X_j)$, where
$X_j$ is subject to the Gaussian distribution with variance $v'$ and mean $0$.
Using the conventional random coding of rate $R_{v'}$
with respect to the above Gaussian distribution, 
we generate a code. 
Here, the choice of the code is a part of the preshared information between Alice and Bob.
When Alice encodes the message $M_n$ via the preshared code, 
using the preshared code, Bob decodes message $M_n$.
However, as is discussed in the paper \cite{Wang},
in this case, Willie cannot distinguish the received signals from 
the Gaussian distribution with variance $v_W+b^2 v'$ and mean $0$, which is a special case of the silent case. 
In this way, this method achieves the covert transmission of $O(n)$ bits.
Although this method has a higher covert transmission speed than our method,
it has the following problem.
When the decoder of this method is a maximum likelihood decoder, 
it depends on the value of $a$ while it does not depend on the value of $v_W$.
That is, the above method satisfies the condition (i), but does not satisfy 
the condition (ii);
\begin{description}
\item[(i)] Willie's output of the code simulates 
the Gaussian distribution with variance $v_W+b^2 v'$ and mean $0$.
\item[(ii)] Bob's decoder does not depend on the channel parameters $a$ and $v_B$.
\end{description}
Although the paper \cite{Uni-cont} proposed a universal coding that works for the AWGN channel,
the encoder of \cite{Uni-cont} is generated by a distribution with finite support.
Hence, use of the method \cite{Uni-cont}
satisfies the condition (ii), but does not satisfy the condition (i).

As another idea, we apply the method \cite{Hou} as follows.
The method \cite{Hou} employs the code for wire-tap channel.
Since the paper \cite[Appendix D]{Haya2} proposed a wire-tap code for the AWGN channels,
the method \cite{Hou} satisfies the condition (i)
 when Bob knows the channel parameters of the channel to Bob.
That is, this alternative method does not satisfy the condition (ii).
Fortunately, our code satisfies both conditions, i.e., 
our method is the first method to achieve both conditions (i) and (ii).
In this sense, our code has an advantage over a simple application of 
the methods \cite{Hou,Wang}.

\section{Analysis of Bob's decoding}\Label{S5}
\subsection{Characterization of correct decoding}
The aim of this section is the asymptotic evaluation of 
the probability that 
Bob correctly decodes all messages including 
the detection of the existence of communication from all senders.
In this subsection, we derive a lower bound of this probability.

Bob's receiving signal is written as
\begin{align}
Y_j= N_{B,j}+\sqrt{\frac{t_n}{n}}\sum_{k=1}^{l_n} a_{i_k} (-1)^{M_{i_k}+S_{i_k,j}}.
\end{align}
The variable $\bar{N}_{B,i}:
=\sum_{j=1}^n (-1)^{S_{i,j}} {N}_{B,j}$ 
is a Gaussian variable with average $0$ and variance $n v_B $.
The variables $M_{i_k}+S_{i_k,j}+S_{i,j}$ with 
$j=1, \ldots, n$ and $k=1, \ldots, l_n$ with $i_k\neq i$
are independent binary random variables subject to the uniform distribution.

When Bob focuses on $\bar{Y}_i$ to recover $M_i$, 
only the term related to $M_i$ is of his interest and 
the remaining terms can be considered as noises.
Hence, by using $N(i):=\bar{N}_{B,i}+  \sqrt{\frac{t_n}{n}}
\sum_{k=1,i_k\neq i}^{l_n}  a_{i_k} 
\sum_{j=1}^n (-1)^{M_{i_k}+S_{i_k,j}+S_{i,j}}$,
the term $\bar{Y}_i$ can be rewritten as
\begin{align}
&\bar{Y}_i= \sum_{j=1}^n (-1)^{S_{i,j}} Y_j \nonumber \\
=&\sum_{j=1}^n (-1)^{S_{i,j}} \Big(
{N}_j+\sqrt{\frac{t_n}{n}}\sum_{k=1}^l a_{i_k} (-1)^{M_{i_k}+S_{i_k,j}}
\Big) \nonumber \\
=&\bar{N}_{B,i}+\sqrt{\frac{t_n}{n}}
\sum_{k=1}^{l_n} a_{i_k} \sum_{j=1}^n (-1)^{M_{i_k}+S_{i_k,j}+S_{i,j}}
\nonumber \\
=&\bar{N}_{B,i}+ \sqrt{n t_n}a_i (-1)^{M_i}+\sqrt{\frac{t_n}{n}}
\sum_{k=1,i_k\neq i}^{l_n}  a_{i_k} \sum_{j=1}^n (-1)^{M_{i_k}+S_{i_k,j}+S_{i,j}} \nonumber \\
=&\sqrt{n t_n}a_i (-1)^{M_i}+{N}(i)
\Label{XM3}.
\end{align}
That is, $N(i)$ is considered as a noise.

Assume that ${\cal A}_i$ is silent, i.e., $i \in(\{i_k\}_{k=1}^{l_n})^c $.
Bob's decoding is correct when $|N(i)| <  \sqrt{n t_n} \frac{\underline{a}}{2}$.
That is,
the probability of Bob's correct decoding is 
${\rm Pr}\Big(| N(i)| <  \sqrt{n t_n} \frac{\underline{a}}{2}\Big)$.
See Fig. \ref{FF1C} to illustrate this process.

Assume that ${\cal A}_i$ is active.
Bob's decoding is correct when $M_i=0$ and
$N(i) \le  \sqrt{n t_n} (a_i-\frac{\underline{a}}{2})$.
Also,
Bob's decoding is correct when $M_i=1$ and
$N(i) \ge -  \sqrt{n t_n} (a_i-\frac{\underline{a}}{2})$.
Since $N(i)$ is symmetric, i.e., the distribution of $N(i)$ is the same as 
the distribution of $-N(i)$,
the probability of Bob's correct decoding is 
${\rm Pr}\Big(N(i) \le  \sqrt{n t_n} (a_i-\frac{\underline{a}}{2})\Big)$.

Then, we obtain the following lower bound for 
the probability that 
Bob correctly decodes all messages including 
the detection of the existence of communication from all senders.
\begin{align}
P_n
:=&\prod_{k=1}^{l_n} {\rm Pr} \Big( N(i_k) \le  \sqrt{n t_n} 
(a_{i_k}-\frac{\underline{a}}{2})
\Big) \nonumber \\
&\cdot  
\prod_{ i \in (\{i_k\}_{k=1}^{l_n})^c }
{\rm Pr}\Big(| N(i)| <  \sqrt{n t_n} \frac{\underline{a}}{2}\Big)\Label{AMP}.
\end{align}
Then, 
as shown in Appendix \ref{APP1}, 
we have the following lemma.
\begin{lemma}\Label{LLLA}
When $l$ is given as $l_n=\frac{n}{t_n}$ and the conditions 
\eqref{BZI2} and \eqref{BZI} hold,
we have
\begin{align}
P_n
\to 1,\Label{XMAY}
\end{align}
Since the probability $P_n$ is a non-negative value 
upper bounded by $1$,
its convergence speed to 1 
is evaluated by 
the speed of the convergence of 
$\log P_n$ to $0$.
That is, this convergence is evaluated by
the speed of the convergence of 
$\log (-\log P_n)$ to $- \infty$.
The following expresses an upper bound of 
this convergence speed.
\begin{align}
&\log (-\log P_n )
\le  -\frac{n \underline{a}^2}{8l_n (   v_B+ 
  \overline{a}^2 )}
+ \log m_n +\log 2 +o(1)
.\Label{XMAYS}
\end{align}
The condition \eqref{BZI}
guarantees that this lower bound 
goes to $-\infty$.
\end{lemma}

\subsection{Single sender case}\Label{SSL1}
The single sender case can be evaluated by putting
$ \underline{a}=\overline{a}=a$ and $m_n=l_n$.
In the same way as \eqref{XM3}, we have 
\begin{align}
\bar{Y}_i
=\sqrt{n t_n}a (-1)^{M_i}+{N}(i)\Label{XM43},
\end{align}
where
\begin{align}
{N}(i):=\bar{N}_i+ a
\sqrt{\frac{t_n}{n}}
\sum_{i'=1,i\neq i}^{l_n} \sum_{j=1}^n (-1)^{M_{i'}+S_{i',j}+S_{i,j}}.
\end{align}
Since ${N}(i)$ is symmetric,
in the same way as \eqref{AMP},
the probability of correct decoding is
lower bounded as
\begin{align}
\prod_{k=1}^{l_n} {\rm Pr} \Big( N(i) < \sqrt{n t_n} a
\Big) \Label{AMP2}.
\end{align}
Then, as a special case of Lemma \ref{LLLA}, 
we have the following lemma.
\begin{lemma}\Label{LLLA2}
When $l$ is given as $l_n=\frac{n}{t_n}$ and the conditions 
\eqref{BZI2} and \eqref{ALZ2-A} hold,
we have
\begin{align}
\prod_{k=1}^{l_n} {\rm Pr} \Big( N(i) < \sqrt{n t_n} a
\Big)
\to 1.\Label{XMAYC}
\end{align}
\end{lemma}

\section{Willie's detection}\Label{S5-2}
\subsection{Formulation}\Label{S5-2A}
Since the random access channel model contains the point-to-point channel 
as a special case with
$ \underline{b}=\overline{b}=b$ and $m_n=l_n$,
we discuss the random access channel model in the following.
Under the encoder \eqref{NJ}, Willie's receiving signal is written as
\begin{align}
Z_j=N_{W,j}+\frac{1}{\sqrt{n}}\sum_{k=1}^{l_n} b_{i_k}  (-1)^{M_{i_k}+S_{i_k,j}}.
\Label{XM5}
\end{align}
We denote the distribution for $Z_j$ by $P_{Z_j}$
and denote the joint distribution for $Z=(Z_1, \ldots, Z_n)$ by 
$P_Z$.
$P_{Z|M=m}$ expresses the conditional distribution under the condition
$M(:=(M_1, \ldots, M_{l_n})) =m(:=(m_1, \ldots, m_{l_n}))$.

We denote the Gaussian distribution with average $0$ and variance $v$ by
$G_v$.
When all senders are silent, 
Willie's observation 
$(Z_1,\ldots, Z_n) $ is subject to the distribution 
$G_{v_W}^{n}$.
Also, we assume that 
Willie does not know the exact value of the variance $v_W$ of his receiving device
while Willie knows its rough value
because a part of the noise $N_{W,j}$ is generated out of Willie's device, 
which can be considered as a background noise.
Then, we denote the set of possible variances with the silent case
by ${\cal V}$. 
In the following, we assume that ${\cal V}$ is an open set of $\mathbb{R}$.

In this case, to cover Willie's advantageous scenario, 
we assume that Willie knows the secret message $M$, but he does not know
the binary symbols $\{S_{i,j}\}$.
That is, we show that Willie cannot detect the existence of the communication 
even though he knows the secret message $M$.
When no communication is made, 
the joint distribution of Willie' receiving signal $Z$ and the secret message 
belongs to the set $\{ G_{v_W}^n \times P_M| v_W \in {\cal V} \}$.

It is known that the statistical distinguishablity is characterized by the variational distance as follows,
where we denote the variational distance between $P$ and $Q$ by $d_V(P,Q)$.
Given a method ${\cal T}$ to distinguish $P$ and $Q$,
we denote 
the probability being incorrectly deciding the distribution to be $P$ while the true is $Q$, by 
$\epsilon_1({\cal T})$,
and 
the probability being incorrectly deciding the distribution to be $Q$ while the true is $P$, by 
$\epsilon_2({\cal T})$.
The sum of 
$\epsilon_1({\cal T})$ and $\epsilon_2({\cal T})$
is evaluated as \footnote{This fact is well known
in the community of quantum information.
For example, the reader might see the reference
\cite[Section 3.2]{Haya}}.
\begin{align}
\epsilon_1({\cal T})+\epsilon_2({\cal T})
\ge 1-d_V(P,Q).
\end{align}

Hence, the distinguishability between 
the real distribution $P_{Z,M}$ of making the communication 
and the case with no communication is measured by 
the minimum
$\min_{{v} \in {\cal V}} d_V(P_{Z,M} , G_{{v}}^n \times P_M )$,
which shows the ability of distinguishing the following two hypotheses.
One is the hypothesis that the true distribution is $P_{Z,M}$, which corresponds to the case of
making the communication.
The other one is the hypothesis that 
the true distribution belongs to the set $\{ G_{v_W}^n \times P_M| v_W \in {\cal V} \}$, 
which corresponds to the case of no communication.
Hence, when our covertness measure
$ \min_{{v} \in {\cal V}} d_V(P_{Z,M} , G_{{v}}^n \times P_M )$ is sufficiently small with 
any $l$ active senders,
the situation with any $l$ active senders
cannot be distinguished with 
the situation with no communication, 
i.e., Willie cannot detect any active user.

\begin{theorem}\Label{TH2B}
Assume that $l$ is given as $l_n=\frac{n}{t_n}$ to satisfy the condition \eqref{NCA}.
Also, we assume that $v_W$ belongs to ${\cal V}$,
Then, 
our covertness measure goes to zero as
under the protocol presented in Section \ref{S-pro}
\begin{align}
 \min_{{v} \in {\cal V}} d_V(P_{Z,M} , G_{{v}}^n \times P_M )\to 0
 \Label{ACCY}.
\end{align}
\end{theorem}
The combination of Lemma \ref{LLLA}
and Theorem \ref{TH2B}
leads Theorem \ref{TH2}.

Further, under the case with equal fading in Willie's detection, 
the condition in Theorem \ref{TH2B} can be relaxed as follows.
\begin{theorem}\Label{TH2BE}
Assume that 
$ \underline{b}=\overline{b}=b$ and
$l$ is given as $l_n=\frac{n}{t_n}$ to satisfy the condition
\begin{align}
\frac{n}{l_n^2 },
\frac{l_n^2 }{n^3} 
&\to 0
\Label{NCAE}.
\end{align}
Also, we assume that $v_W$ belongs to ${\cal V}$,
Then, 
our covertness measure goes to zero 
under the protocol presented in Section \ref{S-pro}
as
\begin{align}
 \min_{{v} \in {\cal V}} d_V(P_{Z,M} , G_{{v}}^n \times P_M )\to 0
 \Label{ACCYE}.
\end{align}
\end{theorem}

When $l_n$ is linear with $n$, 
the condition \eqref{NCA}
does not hold, but
the condition \eqref{NCAE} does hold.
In this sense, Theorem \ref{TH2BE} has a weaker condition 
for $l_n$ than Theorem \ref{TH2B}.
The case with equal fading in Willie's detection, i.e., the case with 
$ \underline{b}=\overline{b}=b$, covers 
the case with the point-to-point channel.
Thus, due to Theorem \ref{TH2BE}, 
one sender, Alice, can send $l_n$ bits to Bob 
in one channel coherence time securely with covertness to Willie.
the conditions 
\eqref{BZI2} and \eqref{BZI} 
imply the condition \eqref{NCAE},
the combination of Lemma \ref{LLLA2}
and Theorem \ref{TH2BE}
leads Theorem \ref{TH1}.

\subsection{Useful formula for our covertness measure}
As a preparation of our proofs of 
Theorems \ref{TH2B}
and \ref{TH2BE},
we prepare a useful formula for our covertness measure, 
i.e., 
the minimum variational distance as follows.
\begin{align}
& \min_{{v} \in {\cal V}} d_V(P_{Z,M} , G_{{v}}^n \times P_M )\nonumber \\
\le &
\min_{{v} \in {\cal V}} d_V( D(P_{Z,M}, P_Z \times P_M)+
d_V( P_Z  \times P_M, G_{{v}}^n \times P_M) \nonumber \\
= &
d_V(P_{Z,M}, P_Z \times P_M ) 
+\min_{{v} \in {\cal V}} d_V( P_Z  , G_{{v}}^n).
\Label{ACC}
\end{align}
The first term $d_V(P_{Z,M}, P_Z \times P_M )$ expresses the secrecy of the message,
and the second term
$\min_{{v} \in {\cal V}}  d_V( P_Z  , G_{{v}}^n)$
expresses the possibility that Willie detects the existence of the communication.

The variables $(M_i+S_{i_k,j})_{k,j}$ are independently subject to the binary uniform distribution 
under the condition $M=m$.
The variables $Z=(Z_1, \ldots, Z_n)$ do not depend on $m$.
That is, we have $P_{Z|M=m}=P_{Z}$.
Hence, Willie has no information for the message $M=m$, i.e., 
\begin{align}
d_V(P_{Z,M}, P_Z \times P_M )=0.\Label{XMR}
\end{align}

In the following, we discuss 
$\min_{{v} \in {\cal V}}  d_V( P_Z , G_{{v}}^n)$.
We choose $v':= \frac{1}{n}\sum_{k=1}^l b_{i_k}^2$.
When $c=\frac{l}{n}$ is sufficiently small,
$v'$ is small so that
$v_W+v' $ belongs to ${\cal V}$ because 
$v_W \in {\cal V}$ and ${\cal V}$ is an open set.
Hence, we have
\begin{align}
\min_{v \in {\cal V}}  d_V( P_Z  , G_{v}^n)
\le d_V( P_Z , G_{v_W+v'}^n).\Label{ZZL}
\end{align}
Combining \eqref{ACC}, \eqref{XMR}, and \eqref{ZZL}, we obtain
\begin{align}
\min_{{v} \in {\cal V}} d_V(P_{Z,M} , G_{{v}}^n \times P_M )
\le d_V( P_Z , G_{v_W+v'}^n).
\end{align}

Applying the Pinsker inequality
$d_V(P,Q)^2 \le \frac{1}{2} D(P\|Q)$, where $D(P\|Q):= \int (\log P(x)-\log Q(x))  P(x) dx$,
we have $
d_V( P_Z , G_{v_W+v'}^n)^2 \le \frac{1}{2} 
D( P_{Z_1,\ldots, Z_n}\| G_{v_E+v'}^{n} )$.
Hence, it is sufficient to show that 
the relative entropy $D( P_{Z_1,\ldots, Z_n}\| G_{v_E+v'}^{n} )$
between the joint distribution $P_{Z_1,\ldots, Z_n}$
of the random variables $Z_1,\ldots, Z_n$ and 
the $n$-fold Gaussian distribution $G_{v_E+v'}^{n}$
is sufficiently small.
Combining \eqref{ACC}, \eqref{XMR}, and the above application of the Pinsker inequality,
we have  
\begin{align}
\min_{{v} \in {\cal V}} d_V(P_{Z,M} , G_{{v}}^n \times P_M )\le 
\sqrt{\frac{1}{2} 
D( P_{Z_1,\ldots, Z_n}\| G_{v_E+v'}^{n} )}.\Label{ZMT}
\end{align}
Therefore, 
the remaining part evaluates $D( P_{Z_1,\ldots, Z_n}\| G_{v_E+v'}^{n} )$.

\subsection{Analysis of case with equal fading}\Label{S5C}
Now, we show Theorem \ref{TH2BE}
by using the result in \cite{IID}.
We assume that 
$ \underline{b}=\overline{b}=b$.
To state the result by \cite{IID},
we define $\chi^2$ distance $\chi^2(P,Q)$ as
\begin{align}
\chi^2(P,Q):= \int \frac{(p(x)-q(x))^2}{q(x)} dx,
\end{align}
where $p$ and $q$ are probability density functions of the distributions
$P$ and $Q$.
We define Renyi divergence of order $2$,
$D_2(P\|Q)$ as
 \begin{align}
D_2(P\|Q):= \log \int \frac{p(x)^2}{q(x)} dx.
\end{align}
We have
\begin{align}
D(P\|Q)\le D_2(P\|Q) \le \log  (1+\chi^2(P,Q)) \le \chi^2(P,Q).
\Label{IMP}
\end{align}

\begin{proposition}[\protect{\cite[(1.3)]{IID}}]\Label{Pro1}
$U_1, \ldots, U_n$ are $n$ independent and identical distributed variables
with average $0$ and variance $v_W$.
The distribution of $U_i$ is absolutely continuous and has the probability 
density function $P_U(u)$.
$P_U(u)$ is symmetric, i.e., $P_U(u)=P_U(-u)$.
Define 
\begin{align}
{\cal U}_n:= \sum_{i=1}^n \frac{1}{\sqrt{n}} U_i.
\end{align}
Then, we have
\begin{align}
\chi^2( P_{{\cal U}_n}, G_1 )
=\frac{(\alpha_4-3)^2}{24n^2}+O(\frac{1}{n^3}),
\end{align}
where $\alpha_4:=\mathbb{E}[G_1^4]$.
\end{proposition}

Since $v'$ is simplified as $v'= \frac{l_n b^2}{n} =\frac{b^2}{t_n} $, we have
\begin{align}
&\alpha_4(j):=\mathbb{E}\Big[\Big(\frac{1}{\sqrt{v+v'}} Z_j\Big)^4\Big]
=\frac{{v'}^2+6 v_W v'+3v^2}{(v_W +v')^2} \nonumber \\
=&\frac{{v'}^2+6 v_W v'+3v^2-3(v_W +v')^2}{(v_W +v')^2}+3 
=\frac{-2(\frac{ b^2}{t_n})^2}{(v_W +\frac{b^2}{t_n})^2}+3\Label{ACO}.
\end{align}
Since $ \underline{b}=\overline{b}=b$,
we have $b_{i_k}=b$ for all $k$, Proposition \ref{Pro1} guarantees that
\begin{align}
&\chi^2( P_{Z_j}, G_{v_W +v'} )
=\chi^2( P_{ \frac{1}{\sqrt{v+v'}} Z_j}, G_{1} )\nonumber \\
=&\frac{\big(\alpha_4(j)-3\big)^2}{24 l^2}
+O(\frac{1}{l_n^3}) \nonumber \\
\stackrel{(a)}{=} 
&\frac{4(\frac{b^2}{t_n})^4}{24 l_n^2 (v_W +\frac{b^2}{t_n})^4}
+O(\frac{1}{l_n^3}) 
=\frac{b^8 }{6 n^2t_n^2 (v_W  +b^2 t_n^{-1})^4}
+O(\frac{t_n^3}{n^3}) ,
\end{align}
where $(a)$ follows from \eqref{ACO}.
Hence, we have
\begin{align}
D( P_{Z_j}\| G_{v_W +v'} )
\le  \frac{b^8 }{6 n^2t_n^2 (v_W  +b^2 t_n^{-1})^4}
+O(\frac{t_n^3}{n^3}) .
\end{align}
Thus,
\begin{align}
& D( P_{Z_1,\ldots, Z_n}\| G_{v_W +v'}^{n} )
\le n\Big( 
\frac{b^8 }{6 n^2t_n^2 (v_W  +b^2 t_n^{-1})^4}
+O(\frac{t_n^3}{n^3}) \Big) \nonumber \\
=&
\frac{b^8 }{6 n t_n^2 (v_W  +b^2 t_n^{-1})^4}
+O(\frac{t_n^3}{n^2}) .
\Label{ACS}
\end{align}
The condition \eqref{NCAE} guarantees that
$\frac{b^8 }{6 n t_n^2 (v_W  +b^2 t_n^{-1})^4}$
and $\frac{t_n^3}{n^2}$ go to zero, which implies
that the term $D( P_{Z_1,\ldots, Z_n}\| G_{v_W +v'}^{n} )$ goes to zero.
Hence, combining \eqref{ZMT},
we obtain Theorem \ref{TH2BE}.

\subsection{Analysis of case with unequal fading}\Label{S5B}
Now, we show Theorem \ref{TH2B}.
We consider the case with unequal fading in Willie's detection.
For the analysis of this case, we have the following theorem.
\begin{theorem}\Label{TUI}
The relative entropy $D( P_{Z_1,\ldots, Z_n}\| G_{ v_W +v'}^{n} )$ is evaluated as
\begin{align}
& D( P_{Z_1,\ldots, Z_n}\| G_{ v_W +v'}^{n} )\nonumber \\
\le &
\frac{\frac{1}{ n{v'}^{2}} \sum_{k=1}^{l_n} b_{i_k}^4
\Big(\frac{(v_W +v')(e^{\frac{v'}{v_W +2v'}}+e^{-\frac{v'}{v_W }} )}{2\sqrt{v_W (v_W +2v')}}-1\Big)}
{\frac{e^{-\frac{v'+v_W }{2v_W }}}{2}+(1-\frac{e^{-\frac{v'+v_W }{2v_W }}}{2})\frac{1}{ n^2{v'}^{2}} \sum_{k=1}^{l_n} b_{i_k}^4}.
\Label{NM}
\end{align}
\end{theorem}

Since the proof of this theorem is very long, Appendix \ref{APC} proves it 
by using the result in \cite{non-IID2}, which employs the Poincar\'{e} constant.

We choose $c_n:= \frac{l_n}{n}=\frac{1}{t_n}$.
We have $\sum_{k=1}^{l_n} b_{i_k}^4 \le {l_n} \overline{b}^4
= c_n\overline{b}^4 n$
and $v'\le \frac{l_n \overline{b}^2}{n}=c_n \overline{b}^2$.
Since the condition \eqref{NCA}
by using \eqref{XMD},
the RHS of \eqref{NM} is evaluated as
\begin{align}
&(\hbox{ RHS of }\eqref{NM})\nonumber \\
\le &
\frac{\sum_{k=1}^{l_n} b_{i_k}^4 }{n {v'}^2}
\frac{1}{\frac{e^{-\frac{v'+v_W }{2v_W }}}{2}} 
\Big(\frac{1}{6}\Big(\frac{v' }{v_W }\Big)^3
+ O\Big( \Big(\frac{v' }{v_W }\Big)^4\Big)\Big) \nonumber \\
= &
\Big(\sum_{k=1}^{l_n} b_{i_k}^4 \Big)
\frac{e^{1/2}}{3}\Big(\frac{v' }{n v_W^3}\Big)
\Big(1+O\Big(\frac{v' }{v_W }\Big)\Big)
+ O\Big( \Big(\frac{v' }{v_W }\Big)^4\Big) \nonumber \\
\le &
c_n \overline{b}^4
\frac{e^{1/2}}{3}
\Big(\frac{c \overline{b}^2}{v_W^3}\Big)(1+O(c_n))+ O( c_n^4) \nonumber \\
= &
\frac{c_n^2 e^{1/2}}{3}
\Big(\frac{\overline{b}^2}{v_W }\Big)^3+ O( c_n^3) .
\end{align}

Since the condition \eqref{NCA}
guarantee that $c_n\to 0$,
we have 
\begin{align}
D( P_{Z_1,\ldots, Z_n}\| G_{ v_W +v'}^{n} )
\to 0.
\end{align}
Hence, combining \eqref{ZMT},
we complete the proof of Theorem \ref{TH2B}.

Here, we compare the evaluation of \eqref{NM} with
the evaluation of \eqref{ACS}.
Eq. \eqref{NM} has the order $O(\frac{1}{t_n^2})=O(\frac{1}{\log n^2})$ for $n$,
and Eq. \eqref{ACS} has the order $O(\frac{1}{n t_n^2})=O(\frac{1}{n \log n^2})$ for $n$.
Hence, Eq. \eqref{ACS} 
results in a much better covertness evaluation than 
Eq. \eqref{NM}.
Therefore, combining this discussion and Lemma \ref{LLLA},
we obtain Theorem \ref{TH2}.

\section{Discussion and open problems}\Label{S6}
We have discussed covert communication assuming BPSK over AWGN channels.
Our results are composed of two contributions:
covert communications for the random access channel
and point-to-point communication.
In the former case, we assume that the sender can choose the power to 
a certain value that is fixed during 
a coding block length.
Our encoder and our decoder in both settings are quite simple.
Thus, the proposed methods are easily implementable 
being the main complexity the problem of key sharing, as in traditional spread spectrum. 
However, we have not described the detail  
of the implementation of our method based on spread-spectrum principle,
which is needed for practical application.
Its detailed description is an important future study.

In our method, we assume that Willie does not have perfect knowledge on the channel parameter.
That is, using this lack of Willie's knowledge, our method improves the transmission speed
over existing methods \cite{Yan,Wang2,Zhang} in the case of AWGN channel.
The key idea for our method is 
convergence of the distribution of the weighted sample mean of $n$ independent random variables 
to a Gaussian distribution, which is related to a kind of central limit theorem \cite{non-IID}.
This method depends on the fact that our output distribution is a Gaussian distribution.
That is, it is impossible to extend this method to another channel model.
Due to this reason, we also assume that Bob 
does not have perfect knowledge on the channel parameter.
Our code
is shown to guarantee that Bob reliably recovers the message under this condition.
That is, our code is a universal code in this sense unlike the recent papers \cite{Lee,He,Sobers,Soltani,Goeckel,Hu,Shmuel,Xiong}.

In addition, since our covertness analysis needs the evaluation of the variational distance,
we need more precise evaluation than the variant of central limit theorem \cite{non-IID}.
To resolve this problem, we have employed the results from
the papers \cite{IID,non-IID2,gap}. 
As discussed in Section \ref{S5C},
the result in \cite{IID} has been used for the analysis on the case with equal fading.
For the general case with unequal fading,
as discussed in Section \ref{S5B},
the analysis on the papers \cite{non-IID2,gap} employs
Poincar\'{e} constant \cite{Bobkov,Bobkov2,gap,non-IID2}.
Although Section \ref{S5B} analyzes the case with unequal fading,
the evaluation in Section \ref{S5C} is tighter than in Section \ref{S5B}.
Hence, 
the above analysis has better evaluation for the ability of Willie's detection of the existence of the communication
than the use of evaluation in Section \ref{S5B}.
Hence, Eq. \eqref{ACS} suggests a possibility to improve the evaluation 
\eqref{NM}.
For this improvement, it is needed to extend Proposition \ref{Pro1} 
to the case \eqref{unequal}, i.e., the case with unequal weighted sum. 
It is an interesting a future study.


Further, we have not proved the converse part for the transmission length for our covert communication model.
Since our code construction is based on an elementary idea,
there is a possibility to improve the transmission speed 
of our method. 
It is another future problem to derive the asymptotically tight transmission speed 
under both settings.
In the relation to this topic, to state the advantage of our method for 
the point-to-point communication, we have pointed that
existing methods \cite{Hou,Wang} need perfect knowledge for channel parameters.
That is, our model 
requires a code to satisfy the two conditions (i) and (ii) defined in Section \ref{S3-3}
simultaneously. 
Although our code satisfies both conditions,
we have not shown that no code with transmission length $O(n)$ 
satisfies both conditions in the point-to-point communication.
In fact, in our protocol for the point-to-point communication,
our code is essentially 
constructed with bit-by-bit communication.
Clearly, this method is not efficient when covertness is not discussed.
Therefore, 
it is an interesting point whether or not our protocol can be improved by a more efficient method in the point-to-point scenario.
It is another open problem to clarify this issue.

Also, our method needs pre-shared secret of $O(n^2)$ bits.
Since this is larger than the size of transmitted message,
it is better to reduce this size 
while keeping our transmission rate.
One possibility for this solution is choosing the matrix $S_{i,j}$ to be a Toeplitz matrix.
While this choice reduces the size of pre-shared secret to $O(n)$,
it changes the stochastic behaviors of $Y_j$ and $Z_j$.
It is another future study to evaluate the performance of 
this modified code.
Finally, it is an interesting future study to apply the theoretical limits obtained in this work to a practical engineering setting with realistic fading channel parameters and evaluate the performance with practical BPSK symbol acquisition and realistic channel dynamics using metrics such as covert security outage probability.

\appendices

\section{Proof of Lemma \ref{LLLA}
}\Label{APP1}
To evaluate the above quantity, we introduce two kinds of cummulant generating  functions $\phi_{b}$ and $\phi_{g,v}$ as
\begin{align}
\phi_{b}(s):= \log \frac{e^{s}+e^{-s}}{2}, \quad 
\phi_{g,v}:= \frac{vs^2}{2}.
\end{align}
 That is, $\phi_{b}$ is the 
 cummulant generating  function of the variable $(-1)^X$ when $X$ obeys the binary distribution with probability $\frac{1}{2}$,
 and $\phi_{g,v}$ is the cummulant generating  function of Gaussian distribution with average 0 and variance $v$.
Therefore, 
the cummulant generating  function of $N(i)$ is 
$\phi_{n,i}(s):=\phi_{g,n v_B}(s)+\sum_{k=1,i_k\neq i}^{l_n} n \phi_{b}(
a_{i_k} s \sqrt{\frac{ t_n}{n}})$.
We have
\begin{align}
\phi_{n,i}(s)\le \phi_{n}(s):=
\phi_{g,n v_B}(s)+l_n n \phi_{b} \Big(
\overline{a} s \sqrt{\frac{ t_n}{n}}\Big).
\Label{aNX}
\end{align}
When $n$ is large,
\begin{align}
\phi_{n}(s)
&=\frac{s^2}{2} (  n v_B+ l_n  \overline{a}^2 t_n)
+O\Big( nl\Big(\frac{t_n}{n}\Big)^{\frac{3}{2}} s^3 \Big) 
\nonumber \\&
=\frac{s^2}{2}(  n v_B+ l_n  \overline{a}^2 t_n)
+O\Big( l\sqrt{\frac{t_n^3}{n}} s^3\Big)\Label{MM8}.
\end{align}

We assume that ${\cal A}_i$ is silent.
Since the condition \eqref{BZI2} guarantees 
$\frac{ n}{l_n^2}\to 0$,
using \eqref{MM8}, 
we have
\begin{align}
\beta_{n}:=&
 \max_{s}   
\Big(s \sqrt{n t_n} \frac{\underline{a}}{2}-  \phi_{n}(s)\Big) 
\nonumber \\
=&
 \frac{(\sqrt{n t_n} \frac{\underline{a}}{2})^2}{2 (  n v_B+
 l_n \overline{a}^2 t_n)}
+O\Big( l_n\sqrt{\frac{t_n^3}{n}} 
(\frac{\sqrt{t_n}}{\sqrt{n}v_B})^3\Big) \nonumber \\
=&
 \frac{n t_n \underline{a}^2}{8n (   v_B+ 
  \overline{a}^2 )}
+O\Big( \frac{ l_n t_n^3}{n^2 v_B^3}\Big) 
=
 \frac{t_n \underline{a}^2}{8 (   v_B+ 
  \overline{a}^2 )}
+o(1)
\Label{MM9}.
 \end{align}
Thus, under the condition \eqref{BZI}, we have
 \begin{align}
&m_n \exp(-\beta_{n})
=\exp( \log m_n  -\beta_{n})
\to 0. \Label{ZDO1}
 \end{align}
Also, the condition \eqref{BZI} guarantees 
 \begin{align}
 \Big(1 -2\exp ( - \beta_{n} ) \Big)^{\frac{1}{2}\exp ( \beta_{n} )}
 \to \frac{1}{e}.\Label{ZDO2}
\end{align}

Markov inequality implies
\begin{align}
1- {\rm Pr}\Big(| N(i)| <  \sqrt{nt_n} \frac{\underline{a}}{2}\Big)
\le & 2 \exp \Big( - \max_{s}   
\Big(s \sqrt{n t_n} \frac{\underline{a}}{2}-  \phi_{n,i}(s)\Big)\Big) \nonumber \\
\le &
2 \exp ( - \beta_{n})
\Label{MM10}.
\end{align}
Therefore,
\begin{align}
&\prod_{ i \in (\{i_k\}_{k=1}^{l_n})^c }{\rm Pr}\Big(| N(i)| <  \sqrt{n t_n} \frac{\underline{a}}{2}\Big)
\ge \Big(1 -2 \exp ( - \beta_{n} ) \Big)^{m_n-l_n} 
\Label{AXZ}.
\end{align}

We assume that ${\cal A}_i$ is active.
Since we have
\begin{align}
s \sqrt{n t_n} \Big(a_i-\frac{\underline{a}}{2}\Big)-  \phi_{n,i}(s)
\ge s \sqrt{n t_n} \frac{\underline{a}}{2}-  \phi_{n}(s)
\Label{MM11}
\end{align}
for $s \ge 0$, we have
\begin{align}
\max_{s}   
\Big(s \sqrt{n t_n} (a_i-\frac{\underline{a}}{2})-  \phi_{n,i}(s)\Big)\ge \beta_{n}.
\end{align}
Markov inequality implies
\begin{align}
&1-
{\rm Pr}\Big(N(i) <  \sqrt{n t_n} (a_i-\frac{\underline{a}}{2})\Big)
\nonumber \\
\le &
\exp\Big(-\max_{s}   
\Big(s \sqrt{n t_n} (a_i-\frac{\underline{a}}{2})-  \phi_{n,i}(s)\Big)\Big) 
\le \exp (-\beta_{n}).
\end{align}
Therefore, 
we have
\begin{align}
\prod_{k=1}^{l_n} {\rm Pr} \Big( N(i_k) \le  \sqrt{n t_n} 
(a_{i_k}-\frac{\underline{a}}{2})
\Big)
\ge 
(1- \exp (-\beta_{n}))^{l_n} 
. \Label{AM2}
\end{align}
Combining \eqref{AXZ} and \eqref{AM2}, we have
\begin{align}
P_n=&\prod_{k=1}^{l_n} {\rm Pr} \Big( N(i_k) \le  \sqrt{n t_n} 
(a_{i_k}-\frac{\underline{a}}{2})
\Big) \nonumber \\
&\cdot
\prod_{ i \in \{i_k\}_{k=1}^{l_n} }{\rm Pr}\Big(| N(i)| <  \sqrt{n t_n} \frac{\underline{a}}{2}\Big) \nonumber \\
\ge &
(1- \exp (-\beta_{n}))^{l_n} \cdot
\Big(1 -2 \exp ( - \beta_{n} ) \Big)^{m_n-l_n} \nonumber
\\
\ge &\Big(1 -2 \exp ( - \beta_{n} ) \Big)^{m_n} \nonumber \\
=&
\Big(\Big(1 -2 \exp ( - \beta_{n} ) \Big)^{\frac{1}{2}\exp ( \beta_{n} )}\Big)^{2 m_n 
 \exp ( -\beta_{n} )}
\stackrel{(a)}{\to}  1,\Label{NCT}
\end{align}
where $(a)$ follows from \eqref{ZDO1} and \eqref{ZDO2}.
Hence, we obtain \eqref{XMAY}.

In addition, using \eqref{MM9} and \eqref{ZDO2},
we have 
\begin{align}
& \log (-\log P_n ) \nonumber \\
\stackrel{(a)}{\le} & \log \Big(-\log 
\Big(\Big(1 -2 \exp ( - \beta_{n} ) \Big)^{\frac{1}{2}\exp ( \beta_{n} )}\Big)^{2 m_n 
 \exp ( -\beta_{n} )} \Big)\nonumber \\
= & \log 
\Big({2 m_n  \exp ( -\beta_{n} )}
\Big(-\log
\Big(\Big(1 -2 \exp ( - \beta_{n} ) \Big)^{\frac{1}{2}\exp ( \beta_{n} )}\Big)\Big) \Big)\nonumber \\
= & \log 
(2 m_n)  -\beta_{n} 
+\log \Big(- \log
\Big(\Big(1 -2 \exp ( - \beta_{n} ) \Big)^{\frac{1}{2}\exp ( \beta_{n} )}\Big)\Big) \nonumber \\
\stackrel{(b)}{=}& -\frac{n \underline{a}^2}{8l_n (   v_B+ 
  \overline{a}^2 )}
+ \log m_n 
+\log 2+o(1)+\log (1+o(1)) \nonumber \\
=& -\frac{n \underline{a}^2}{8l_n (   v_B+ 
  \overline{a}^2 )}
+ \log m_n 
+\log 2+o(1),
\end{align}
where
$(a)$ follows from \eqref{NCT}
and
$(b)$ follows from \eqref{MM9} and \eqref{ZDO2}.
This relation implies \eqref{XMAYS}.


\section{Preparation for proof of Theorem \ref{TUI}}
For our proof of  Theorem \ref{TUI}, we make several preparations in this subsection.
First, we introduce the Poincar\'{e} constant.
Assume that the random variable $H$ is subject to a distribution $P$ on $\mathbb{R}$.
We define the Poincar\'{e} constant $C(P)$ for  the distribution $P$ as
\begin{align}
C(P) := \inf_{f: \hbox{smooth on } \mathbb{R} }
\frac{ E[ (f'(H))^2 ]}{V[ f(H) ]}.
\end{align}
The value $C(P)$ will be used in \eqref{XMZ}.
For example, it is known that 
the Poincar\'{e} constant $C(G_v)$ is calculated as \cite{Ledoux,Gross}
\begin{align}
C(G_v)=\frac{1}{v}.
\Label{XCV}
\end{align}

Using the Poincar\'{e} constant, we have the following proposition.
\begin{proposition}[\protect{\cite[Theorem 1]{non-IID2}}]\Label{Pro1B}
$U_1, \ldots, U_n$ are $n$ independent and identical distributed variables
with average $0$ and variance $1$.
The distribution of $U_i$ is absolutely continuous and has the probability 
density function $P_U(u)$.
Define 
\begin{align}
{\cal U}_n:= \sum_{i=1}^n \alpha_i U_i, \quad \Label{unequal}
L(\alpha):= \sum_{i=1}^n \alpha_i^4 ,
\end{align}
where $\sum_{i=1}^n \alpha_i^2=1$.
Then, we have
\begin{align}
D( P_{{\cal U}_n}\| G_1 )
\le 
\frac{L(a)}{\frac{C(p)}{2}+(1-\frac{C(p)}{2})L(a)}
D( P_U\| G_1 )\Label{XMZ}.
\end{align}
\end{proposition}


As another preparation, 
we introduce the probability density function 
$P_{v,v'}$ of the random variable 
$\sqrt{\frac{v'}{v'+v}} \Big( (-1)^{X} + 
\sqrt{\frac{v}{v'}} N \Big)$,
where $X$ is the binary variable subject to the uniform distribution
and $N$ is the standard Gaussian variable. 

The Poincar\'{e} constant of $P_{v,v'}$ is evaluated as follows.
\begin{lemma}\Label{CMR}
We have
\begin{align}
C(P_{v,v'})\ge  e^{-\frac{v'+v}{2v}}.\Label{ACT1}
\end{align}
\end{lemma}

Also, we evaluate the relative entropy between $P_{v,v'}$ and the Gaussian distribution in
the following lemma.
\begin{lemma}\Label{XL1}
\begin{align}
D( P_{v,v'}\| G_1 ) \le \frac{(v+v')(e^{\frac{v'}{v+2v'}}+e^{-\frac{v'}{v}} )
}{2\sqrt{v(v+2v')}}-1.\Label{XMP}
\end{align}
In addition, when $\frac{v'}{v}$ is small, we have
\begin{align}
\frac{(v+v')(e^{\frac{v'}{v+2v'}}+e^{-\frac{v'}{v}} )
}{2\sqrt{v(v+2v')}}-1
= \frac{1}{6}\Big(\frac{v'}{v}\Big)^3
+O\Big( \Big(\frac{v'}{v}\Big)^4\Big).
\Label{XMD}
\end{align}
\end{lemma}
The proofs of Lemmas \ref{CMR} and \ref{XL1} will be given in
Appendices \ref{SubA} and \ref{SubB}, respectively.

\section{Proof of Theorem \ref{TUI}}\Label{APC}
In this subsection, we show Theorem \ref{TUI} by using the above preparation.
We prepare $l_n$ independent standard Gaussian variables $N_1,\ldots, N_{l_n}$.
We have
\begin{align}
Z_j=
&N_{W,j}+\frac{1}{\sqrt{n}}\sum_{k=1}^{l_n} b_{i_k}  (-1)^{M_i+S_{i_k,j}} \nonumber \\
=&\sum_{k=1}^{l_n} \frac{b_{i_k}}{\sqrt{n}} 
\Big( (-1)^{M_i+S_{i_k,j}} + 
\sqrt{\frac{v_W}{v'}}
N_k \Big).
\end{align}

We apply Proposition \ref{Pro1B} to the case
$n={l_n}$,
$i=k$,
$\alpha_k=\frac{b_{i_k}}{\sqrt{n v'}}$,
$U_k= 
\sqrt{\frac{v'}{v'+v_W}} \Big( (-1)^{M_i+S_{i_k,j}} + 
\sqrt{\frac{v_W}{v'}}N_k \Big)$.
Then, we have
\begin{align}
\sum_{k=1}^{l_n} \alpha_k^4 =& \frac{1}{ n^2{v'}^{2}} 
\sum_{k=1}^{l_n} b_{i_k}^4.
\end{align}
Since $Z_j = \sqrt{v_W+v'}\big(\sum_{k=1}^{l_n} \alpha_k S_k \big) $,
applying Proposition \ref{Pro1B},
we have
\begin{align}
& D( P_{Z_j}\| G_{v_W+v'} )= D( P_{\sum_{k=1}^{l_n} \alpha_k S_k}\| G_{1} ) \nonumber \\
\le  & 
\frac{\frac{1}{ n^2{v'}^{2}} \sum_{k=1}^{l_n} b_{i_k}^4}
{\frac{C(P_{v_W,v'})}{2}+
\big(1-\frac{C(P_{v_W,v'})}{2}\big)\frac{1}{ n^2{v'}^{2}} 
\sum_{k=1}^{l_n} b_{i_k}^4}
D( P_{v_W,v'}\| G_1 ) 
\nonumber \\
\stackrel{(a)}{\le} & 
\frac{\frac{1}{ n^2{v'}^{2}} \sum_{k=1}^{l_n} b_{i_k}^4\Big(\frac{(v_W+v')(e^{\frac{v'}{v_W+2v'}}+e^{-\frac{v'}{v_W}} )}{2\sqrt{v_W(v_W+2v')}}-1\Big)
}
{\frac{C(P_{v_W,v'})}{2}+
\big(1-\frac{C(P_{v_W,v'})}{2}\big)\frac{1}{ n^2{v'}^{2}} 
\sum_{k=1}^{l_n} b_{i_k}^4}
\nonumber \\
\stackrel{(b)}{\le}  & 
\frac{\frac{1}{ n^2{v'}^{2}} \sum_{k=1}^{l_n} b_{i_k}^4\Big(\frac{(v_W+v')(e^{\frac{v'}{v_W+2v'}}+e^{-\frac{v'}{v_W}} )}{2\sqrt{v_W(v_W+2v')}}-1\Big)
}
{\frac{e^{-\frac{v'+v_W}{2v_W}}}{2}
+\big(1-\frac{e^{-\frac{v'+v_W}{2v_W}}}{2}\big)\frac{1}{ n^2{v'}^{2}} 
\sum_{k=1}^{l_n} b_{i_k}^4},
\Label{NM3}
\end{align}
\begin{figure*}[!t]
\begin{align}
&\chi^2(P_{v,v'},G_1) +1 
=
\frac{(v+v')(e^{\frac{v'}{v+2v'}}+e^{-\frac{v'}{v}} )}{2\sqrt{v(v+2v')}}\nonumber \\
=&
\frac{v (1+\frac{v'}{v})(2 +\frac{v'}{v+2v'}-\frac{v'}{v} 
+\frac{1}{2}(\frac{v'}{v+2v'})^2
+\frac{1}{6}(\frac{v'}{v+2v'})^3
+\frac{1}{2}(\frac{v'}{v})^2
+\frac{1}{6}(\frac{v'}{v})^3
+O( (\frac{v'}{v})^4)))
}{2v} \nonumber \\ &\cdot 
\Big(1- \frac{v'}{v} 
+\frac{3}{2} (\frac{v'}{v})^2  
-\frac{5}{2} (\frac{v'}{v})^3  
+O\Big( \frac{1}{v}\Big(\frac{v'}{v}\Big)^4\Big) \Big)\nonumber \\
=&
\frac{ (1+\frac{v'}{v})(2 +\frac{v'}{v} (1-2\frac{v'}{v}+(2\frac{v'}{v})^2
)-\frac{v'}{v} 
+\frac{1}{2}(\frac{v'}{v})^2(1-4\frac{v'}{v})
+\frac{1}{6}(\frac{v'}{v})^3
+\frac{1}{2}(\frac{v'}{v})^2
+\frac{1}{6}(\frac{v'}{v})^3
)
}{2}
\nonumber \\ &\cdot 
\Big(1- \frac{v'}{v} 
+\frac{3}{2} (\frac{v'}{v})^2  
-\frac{5}{2} (\frac{v'}{v})^3  
\Big)
+O\Big( \Big(\frac{v'}{v}\Big)^4\Big)\nonumber \\
=&
\frac{ (1+\frac{v'}{v})(2 -(\frac{v'}{v})^2+\frac{7}{3}(\frac{v'}{v})^3
)
}{2}
\cdot \Big(1- \frac{v'}{v} 
+\frac{3}{2} (\frac{v'}{v})^2  
-\frac{5}{2} (\frac{v'}{v})^3  
\Big)
+O\Big( \Big(\frac{v'}{v}\Big)^4\Big)\nonumber \\
=&
 (1+\frac{v'}{v}) \Big(1 -\frac{1}{2}(\frac{v'}{v})^2+\frac{7}{6}(\frac{v'}{v})^3
\Big)
\cdot \Big(1- \frac{v'}{v} 
+\frac{3}{2} (\frac{v'}{v})^2  
-\frac{5}{2} (\frac{v'}{v})^3  
\Big)
+O\Big( \Big(\frac{v'}{v}\Big)^4\Big)\nonumber \\
=&
 \Big(1 -\frac{1}{2}(\frac{v'}{v})^2+\frac{7}{6}(\frac{v'}{v})^3
\Big)
\cdot \Big(
1- \frac{v'}{v} 
+\frac{3}{2} (\frac{v'}{v})^2  
-\frac{5}{2} (\frac{v'}{v})^3  
+\frac{v'}{v} - (\frac{v'}{v})^2 
+\frac{3}{2} (\frac{v'}{v})^3 
\Big)
+O\Big( \Big(\frac{v'}{v}\Big)^4\Big)\nonumber \\
=&
 \Big(1 -\frac{1}{2}\Big(\frac{v'}{v}\Big)^2
 +\frac{7}{6}\Big(\frac{v'}{v}\Big)^3
\Big)
\cdot \Big(
1+\frac{1}{2} \Big(\frac{v'}{v}\Big)^2  
- \Big(\frac{v'}{v}\Big)^3  
\Big)
+O\Big( \Big(\frac{v'}{v}\Big)^4\Big) 
=
1+\frac{1}{6}\Big(\frac{v'}{v}\Big)^3
+O\Big( \Big(\frac{v'}{v}\Big)^4 \Big).\Label{AYE}
\end{align}
\end{figure*}
where $(a)$ and $(b)$ follow from Lemma \ref{XL1} and Lemma \ref{CMR},
respectively.
Therefore,
\begin{align}
& D( P_{Z_1,\ldots, Z_n}\| G_{v_W+v'}^{n} )
=
\sum_{j=1}^n D( P_{Z_j}\| G_{v_W+v'} ) \nonumber \\
\le & n 
\frac{\frac{1}{ n^2{v'}^{2}} \sum_{k=1}^{l_n} b_{i_k}^4\Big(\frac{(v_W+v')(e^{\frac{v'}{v_W+2v'}}+e^{-\frac{v'}{v_W}} )}{2\sqrt{v_W(v_W+2v')}}-1\Big)}
{\frac{e^{-\frac{v'+v_W}{2v_W}}}{2}+
\big(1-\frac{e^{-\frac{v'+v_W}{2v_W}}}{2}\big)
\frac{1}{ n^2{v'}^{2}} \sum_{k=1}^{l_n} b_{i_k}^4}
\nonumber \\
=&
\frac{\frac{1}{ n{v'}^{2}} \sum_{k=1}^{l_n} b_{i_k}^4
\Big(\frac{(v_W+v')(e^{\frac{v'}{v_W+2v'}}+e^{-\frac{v'}{v_W}} )}{2\sqrt{v_W(v_W+2v')}}-1\Big)
}
{\frac{e^{-\frac{v'+v_W}{2v_W}}}{2}+
\big(1-\frac{e^{-\frac{v'+v_W}{2v_W}}}{2}\big)\frac{1}{ n^2{v'}^{2}} \sum_{k=1}^{l_n} b_{i_k}^4}
.
\Label{NM4}
\end{align}

\section{Proof of Lemma \ref{XL1}}\Label{SubB}
The probability density function  
$p_{v,v'}$ of the distribution $P_{v,v'}$ for 
$\sqrt{\frac{v'}{v'+v}} \Big( (-1)^{X} + 
\sqrt{\frac{v}{v'}} N \Big)$ is 
\begin{align}
p_{v,v'}(x)=&
\frac{1}{2} \sqrt{\frac{v'+v}{2\pi v}} e^{-\frac{(v'+v)\big(x- \sqrt{\frac{v'}{v'+v}}\big)^2}{2 v}}
\nonumber \\
&+
\frac{1}{2} \sqrt{\frac{v'+v}{2\pi v}} e^{-\frac{(v'+v)\big(x+ \sqrt{\frac{v'}{v'+v}}\big)^2}{2 v}}.
\end{align}
Then, we have \eqref{AYE} of the top of this page.

Hence, 
\begin{align}
&\chi^2(P_{v,v'},G_1)+1 \nonumber \\
=& 
\int_{-\infty}^\infty 
\Bigg(
\frac{1}{2} \sqrt{\frac{v'+v}{ v}} 
e^{-\frac{(v'+v)\big(x- \sqrt{\frac{v'}{v'+v}}\big)^2}{2 v}
+\frac{x^2}{2}} \nonumber \\&
+
\frac{1}{2} \sqrt{\frac{v'+v}{ v}} 
e^{-\frac{(v'+v)\big(x+ \sqrt{\frac{v'}{v'+v}}\big)^2}{2 v}+\frac{x^2}{2}}
\Bigg)^2
\sqrt{\frac{1}{2\pi }}e^{-\frac{x^2}{2}}
dx \nonumber \\
=& 
\int_{-\infty}^\infty 
\Bigg(
\frac{v'+v}{ 4 v}
e^{-\frac{(v'+v)\big(x- \sqrt{\frac{v'}{v'+v}}\big)^2}{v}+x^2} 
\nonumber \\&
+
\frac{v'+v}{ 4 v}
e^{-\frac{(v'+v)\big(x+ \sqrt{\frac{v'}{v'+v}}\big)^2}{v}+x^2} \nonumber \\
&+
\frac{v'+v}{ 2 v}
e^{-\frac{(v'+v)\big(x- \sqrt{\frac{v'}{v'+v}}\big)^2}{2v}-\frac{(v'+v)\big(x+ \sqrt{\frac{v'}{v'+v}}\big)^2}{2v}+x^2}
\Bigg)
\sqrt{\frac{1}{2\pi }}e^{-\frac{x^2}{2}}
dx \nonumber \\
=& 
\sqrt{\frac{1}{2\pi }}
\int_{-\infty}^\infty 
\Bigg(
\frac{v'+v}{ 4 v}
e^{-\frac{v+2v'}{2v}\big(x- \frac{2\sqrt{v'(v'+v)}}{v+2 v'}\big)^2+\frac{v'}{v+2v'}} \nonumber \\&
+
\frac{v'+v}{ 4 v}
e^{-\frac{v+2v'}{2v}\big(x+ \frac{2\sqrt{v'(v'+v)}}{v+2 v'}\big)^2
+\frac{v'}{v+2v'}}
\nonumber \\&
+
\frac{v'+v}{ 2 v}
e^{-\frac{v+2v'}{2v}x^2-\frac{v'}{v}}
\Bigg)
dx \nonumber \\
=& 
2 \cdot \frac{v'+v}{ 4 v}\sqrt{\frac{v}{v+2v'}}
e^{\frac{v'}{v+2v'}}
+
\frac{v'+v}{ 2 v}\sqrt{\frac{v}{v+2v'}}
e^{-\frac{v'}{v}} \nonumber \\
=&
\frac{v+v'}{2\sqrt{v(v+2v')}}
\big(e^{\frac{v'}{v+2v'}}+e^{-\frac{v'}{v}} \big), 
\end{align}
which implies that
\begin{align}
\chi^2(P_{v,v'},G_1) 
=
\frac{(v+v')(e^{\frac{v'}{v+2v'}}+e^{-\frac{v'}{v}} )
}{2\sqrt{v(v+2v')}}-1.
\Label{IMT}
\end{align}
Combination of \eqref{IMP} and \eqref{IMT} yields \eqref{XMP}. 

When $v'$ is small, $\frac{1}{2\sqrt{v(v+2v')}}
=
\frac{1}{2v}(1- \frac{1}{2}\frac{2v'}{v} 
+\frac{3}{8} (\frac{2v'}{v})^2  
-\frac{5}{16} (\frac{2v'}{v})^3  
+O( \frac{1}{v}(\frac{v'}{v})^4)
=
\frac{1}{2v} \Big(1- \frac{v'}{v} 
+\frac{3}{2} (\frac{v'}{v})^2  
-\frac{5}{2} (\frac{v'}{v})^3  
+O( \frac{1}{v}(\frac{v'}{v})^4) \Big)$.
Hence, using \eqref{AYE} in the top of this page,
we obtain \eqref{XMD}.

\section{Proof of Lemma \ref{CMR}}\Label{SubA}
To show Lemma \ref{CMR}, we prepare the following proposition.
\begin{proposition}[\protect{\cite[Proposition 5]{gap}}]\Label{XMPr}
Two absolutely continuous distributions $\mu$ and $\nu$ satisfy
the inequality
\begin{align}
\frac{\min_{x} \frac{d \nu }{d \mu}(x)}{
\max_{x} \frac{d \nu }{d \mu}(x)}
C_P(\mu)\le C_P(\nu) .
\end{align}
\end{proposition}

We define the function $f(x)$ as 
\begin{align}
f(x):=&\frac{p_{v,v'}(x)}{p_{v,0}(x)}
=\frac{1}{2} \sqrt{\frac{v'+v}{ v}} 
e^{-\frac{(v'+v)}{2v}\big(x- \sqrt{\frac{v'}{v'+v}}\big)^2+\frac{x^2}{2}}
\nonumber \\
&\hspace{10ex}
+
\frac{1}{2} \sqrt{\frac{v'+v}{ v}} 
e^{-\frac{(v'+v)}{2v}\big(x+ \sqrt{\frac{v'}{v'+v}}\big)^2+\frac{x^2}{2}} \nonumber \\
=&
\frac{1}{2} \sqrt{\frac{v'+v}{ v}} 
e^{-\frac{v'}{2v}x^2   + \frac{(v'+v)}{v}\sqrt{\frac{v'}{v'+v}}x -\frac{v'}{2v}}
\nonumber \\&
+
\frac{1}{2} \sqrt{\frac{v'+v}{ v}} 
e^{-\frac{v'}{2v}x^2   - \frac{(v'+v)}{v}\sqrt{\frac{v'}{v'+v}}x -\frac{v'}{2v}}.
\end{align}
For $x \ge 0$, we have
\begin{align}
&f(x)\le  \sqrt{\frac{v'+v}{ v}} 
e^{-\frac{v'}{2v}x^2   + \frac{(v'+v)}{v}\sqrt{\frac{v'}{v'+v}}x -\frac{v'}{2v}}\nonumber \\
=&\sqrt{\frac{v'+v}{ v}} 
e^{-\frac{v'}{2v}\big(x^2   - 2\sqrt{\frac{v'+v}{v'}}x\big) -\frac{v'}{2v}}\nonumber \\
=&\sqrt{\frac{v'+v}{ v}} 
e^{-\frac{v'}{2v}\big(x  - \sqrt{\frac{v'+v}{v'}}\big)^2 +\frac{v+v'}{2v}-\frac{v'}{2v}}\nonumber \\
=&\sqrt{\frac{v'+v}{ v}} 
e^{-\frac{v'}{2v}\big(x  - \sqrt{\frac{v'+v}{v'}}\big)^2 +\frac{1}{2}}
\le \sqrt{\frac{v'+v}{ v}}e^{\frac{1}{2}}.\Label{ACT2}
\end{align}

The derivative of $f$ is calculated as
\begin{align}
&\frac{d f}{dx}(x)\nonumber \\
=&
\frac{1}{2} 
\sqrt{\frac{v'+v}{ v}} 
\Big(-\frac{v'}{v}x + \frac{v'+v}{v}\sqrt{\frac{v'}{v'+v}}\Big)
e^{-\frac{v'}{2v}x^2  + \frac{v'+v}{v}\sqrt{\frac{v'}{v'+v}}x -\frac{v'}{2v}} \nonumber \\
&+
\frac{1}{2} \sqrt{\frac{v'+v}{ v}} 
\Big(-\frac{v'}{v}x - \frac{v'\!+\! v}{v}\sqrt{\frac{v'}{v'\!+\!v}}\Big)
e^{-\frac{v'}{2v}x^2  \! - \! \frac{v'+v}{v}\sqrt{\frac{v'}{v'+v}}x -\frac{v'}{2v}}.
\end{align}
Assume that $x\ge 0$.
The relation $\frac{d f}{dx}(x)=0$ holds if and only if
\begin{align}
&\Big(-\frac{v'}{v}x + \frac{v'+v}{v}\sqrt{\frac{v'}{v'+v}}\Big)
e^{-\frac{v'}{2v}x^2  + \frac{v'+v}{v}\sqrt{\frac{v'}{v'+v}}x -\frac{v'}{2v}} \nonumber \\
=&
\Big(\frac{v'}{v}x + \frac{v'+v}{v}\sqrt{\frac{v'}{v'+v}}\Big)
e^{-\frac{v'}{2v}x^2  - \frac{v'+v}{v}\sqrt{\frac{v'}{v'+v}}x -\frac{v'}{2v}}.
\end{align}
The above relation is equivalent to 
\begin{align}
&e^{2\frac{\sqrt{(v'+v)v'}}{v}x} =
e^{2 \frac{v'+v}{v}\sqrt{\frac{v'}{v'+v}}x }\nonumber \\
=&
\frac{\frac{v'}{v}x + \frac{(v'+v)}{v}\sqrt{\frac{v'}{v'+v}}}{-\frac{v'}{v}x + 
\frac{v'+v}{v}\sqrt{\frac{v'}{v'+v}}}
=
\frac{x + \sqrt{\frac{v'+v}{v'}}}{-x + \sqrt{\frac{v'+v}{v'}}}.
\end{align}
That is,
\begin{align}
&2\frac{\sqrt{(v'+v)v'}}{v}x =
\log \Big(x + \sqrt{\frac{v'+v}{v'}}\Big) 
- \log \Big(-x + \sqrt{\frac{v'+v}{v'}}\Big).
\end{align}
We denote the RHS and the LHS by $g(x)$ and $h(x)$, respectively.
Their derivatives are calculated as
\begin{align}
\frac{d g}{dx}(x)&=2\frac{\sqrt{(v'+v)v'}}{v} \\
\frac{d h}{dx}(x)&=\frac{1}{x + \sqrt{\frac{v'+v}{v'}}}
+\frac{1}{-x + \sqrt{\frac{v'+v}{v'}}}=
\frac{2\sqrt{\frac{v'+v}{v'}}}{-x^2 + {\frac{v'+v}{v'}}}.
\end{align}
Hence, we have
\begin{align}
\frac{\frac{d g}{dx}(0)}{\frac{d h}{dx}(0)}
=\frac{v+v'}{v}>1.
\end{align}

Since $\frac{\frac{d g}{dx}(x)}{\frac{d h}{dx}(x)}$ is monotonically decreasing for $x$,
the solution of 
$\frac{d g}{dx}(x)-\frac{d h}{dx}(x)=0$  in $(0,\frac{v'+v}{v'})$
is only one element $x_0$.
We have $\frac{d g}{dx}(x_0)=\frac{d h}{dx}(x_0)$.
Hence, the solution of 
$g(x)-h(x)=0$  in $(0,\frac{v'+v}{v'})$
is only one element $x_1$.
We have $g(x_1)=h(x_1)$.
We have $g(x)>h(x)$ for $x < x_1$, and
$g(x) <h(x)$ for $x > x_1$.
That is,
$\frac{d f}{dx}(x)>0$ for $x < x_1$, and
$\frac{d f}{dx}(x)<0$ for $\frac{v'+v}{v'}> x > x_1$.
Also, $ \frac{d f}{dx}(x)<0 $ for $x \ge \frac{v'+v}{v'}$.
Hence, we find that
$ \min_{x\ge 0} f(x)$ is realized with $x=0$
and 
$ \max_{x\ge 0} f(x)$ is realized with $x=x_1$.
Thus, we have
\begin{align}
\min_{x\ge 0} f(x)=f(0)=
\sqrt{\frac{v'+v}{ v}} e^{-\frac{v'}{2v}}.
\Label{ACT3}
\end{align}

Combining \eqref{ACT2} and \eqref{ACT3}, we have
\begin{align}
\frac{\min_{x} \frac{d \nu }{d \mu}(x)}{
\max_{x} \frac{d \nu }{d \mu}(x)}
\ge e^{-\frac{v'+v}{2v}}.
\Label{ACT4}
\end{align}
Therefore, \eqref{ACT1} follows from the combination of 
\eqref{XCV}, \eqref{ACT4}, and Proposition \ref{XMPr}.

\end{document}